\documentclass{elsart}
\usepackage{amsfonts,amssymb%
  ,amsmath,bbm,latexsym,epsfig%
  }
\makeatletter
 
\def\diagram{\m@th\leftwidth=\z@ \rightwidth=\z@ \topheight=\z@
\botheight=\z@ \setbox\@picbox\hbox\bgroup}
 
\def\enddiagram{\egroup\wd\@picbox\rightwidth\unitlength
\ht\@picbox\topheight\unitlength \dp\@picbox\botheight\unitlength
\hskip\leftwidth\unitlength\box\@picbox}
 
\def\bfig{\begin{diagram}}
\def\efig{\end{diagram}}
\newcount\wideness \newcount\leftwidth \newcount\rightwidth
\newcount\highness \newcount\topheight \newcount\botheight
 
\def\ratchet#1#2{\ifnum#1<#2 \global #1=#2 \fi}
 
\def\putbox(#1,#2)#3{%
\horsize{\wideness}{#3} \divide\wideness by 2
{\advance\wideness by #1 \ratchet{\rightwidth}{\wideness}}
{\advance\wideness by -#1 \ratchet{\leftwidth}{\wideness}}
\vertsize{\highness}{#3} \divide\highness by 2
{\advance\highness by #2 \ratchet{\topheight}{\highness}}
{\advance\highness by -#2 \ratchet{\botheight}{\highness}}
\put(#1,#2){\makebox(0,0){$#3$}}}
 
\def\putlbox(#1,#2)#3{%
\horsize{\wideness}{#3}
{\advance\wideness by #1 \ratchet{\rightwidth}{\wideness}}
{\ratchet{\leftwidth}{-#1}}
\vertsize{\highness}{#3} \divide\highness by 2
{\advance\highness by #2 \ratchet{\topheight}{\highness}}
{\advance\highness by -#2 \ratchet{\botheight}{\highness}}
\put(#1,#2){\makebox(0,0)[l]{$#3$}}}
 
\def\putrbox(#1,#2)#3{%
\horsize{\wideness}{#3}
{\ratchet{\rightwidth}{#1}}
{\advance\wideness by -#1 \ratchet{\leftwidth}{\wideness}}
\vertsize{\highness}{#3} \divide\highness by 2
{\advance\highness by #2 \ratchet{\topheight}{\highness}}
{\advance\highness by -#2 \ratchet{\botheight}{\highness}}
\put(#1,#2){\makebox(0,0)[r]{$#3$}}}

\def\adjust[#1]{} % For compatibility
 
\newcount \coefa
\newcount \coefb
\newcount \coefc
\newcount\tempcounta
\newcount\tempcountb
\newcount\tempcountc
\newcount\tempcountd
\newcount\xext
\newcount\yext
\newcount\xoff
\newcount\yoff
\newcount\gap%
\newcount\arrowtypea
\newcount\arrowtypeb
\newcount\arrowtypec
\newcount\arrowtyped
\newcount\arrowtypee
\newcount\height
\newcount\width
\newcount\xpos
\newcount\ypos
\newcount\run
\newcount\rise
\newcount\arrowlength
\newcount\halflength
\newcount\arrowtype
\newdimen\tempdimen
\newdimen\xlen
\newdimen\ylen
\newsavebox{\tempboxa}%
\newsavebox{\tempboxb}%
\newsavebox{\tempboxc}%
 
\newdimen\w@dth
 
\def\setw@dth#1#2{\setbox\z@\hbox{\m@th$#1$}\w@dth=\wd\z@
\setbox\@ne\hbox{\m@th$#2$}\ifnum\w@dth<\wd\@ne \w@dth=\wd\@ne \fi
\advance\w@dth by 1.2em}
 
\def\t@^#1_#2{\allowbreak\def\n@one{#1}\def\n@two{#2}\mathrel
{\setw@dth{#1}{#2}
\mathop{\hbox to \w@dth{\rightarrowfill}}\limits
\ifx\n@one\empty\else ^{\box\z@}\fi
\ifx\n@two\empty\else _{\box\@ne}\fi}}
\def\t@@^#1{\@ifnextchar_{\t@^{#1}}{\t@^{#1}_{}}}
\def\to{\@ifnextchar^{\t@@}{\t@@^{}}}
 
\def\t@left^#1_#2{\def\n@one{#1}\def\n@two{#2}\mathrel{\setw@dth{#1}{#2}
\mathop{\hbox to \w@dth{\leftarrowfill}}\limits
\ifx\n@one\empty\else ^{\box\z@}\fi
\ifx\n@two\empty\else _{\box\@ne}\fi}}
\def\t@@left^#1{\@ifnextchar_{\t@left^{#1}}{\t@left^{#1}_{}}}
\def\toleft{\@ifnextchar^{\t@@left}{\t@@left^{}}}
 
\def\two@^#1_#2{\allowbreak
\def\n@one{#1}\def\n@two{#2}\mathrel{\setw@dth{#1}{#2}
\mathop{\vcenter{\lineskip\z@\baselineskip\z@
                 \hbox to \w@dth{\rightarrowfill}%
                 \hbox to \w@dth{\rightarrowfill}}%
       }\limits
\ifx\n@one\empty\else ^{\box\z@}\fi
\ifx\n@two\empty\else _{\box\@ne}\fi}}
\def\tw@@^#1{\@ifnextchar _{\two@^{#1}}{\two@^{#1}_{}}}
\def\two{\@ifnextchar ^{\tw@@}{\tw@@^{}}}
 
\def\tofr@^#1_#2{\def\n@one{#1}\def\n@two{#2}\mathrel{\setw@dth{#1}{#2}
\mathop{\vcenter{\hbox to \w@dth{\rightarrowfill}\kern-1.7ex
                 \hbox to \w@dth{\leftarrowfill}}%
       }\limits
\ifx\n@one\empty\else ^{\box\z@}\fi
\ifx\n@two\empty\else _{\box\@ne}\fi}}
\def\t@fr@^#1{\@ifnextchar_ {\tofr@^{#1}}{\tofr@^{#1}_{}}}
\def\tofro{\@ifnextchar^ {\t@fr@}{\t@fr@^{}}}

\def\mon{\mathop{\m@th\hbox to
      14.6\P@{\lasyb\char'51\hskip-2.1\P@$\arrext$\hss
$\mathord\rightarrow$}}\limits} % width of \epi
\def\leftmono{\mathrel{\m@th\hbox to
14.6\P@{$\mathord\leftarrow$\hss$\arrext$\hskip-2.1\P@\lasyb\char'50%
}}\limits} % width of \epi
\mathchardef\arrext="0200       % amr minus for arrow extension (see \into)

\def\settypes(#1,#2,#3){\arrowtypea#1 \arrowtypeb#2 \arrowtypec#3}
\def\settoheight#1#2{\setbox\@tempboxa\hbox{#2}#1\ht\@tempboxa\relax}%
\def\settodepth#1#2{\setbox\@tempboxa\hbox{#2}#1\dp\@tempboxa\relax}%
\def\settokens`#1`#2`#3`#4`{%
     \def\tokena{#1}\def\tokenb{#2}\def\tokenc{#3}\def\tokend{#4}}
\def\setsqparms[#1`#2`#3`#4;#5`#6]{%
\arrowtypea #1
\arrowtypeb #2
\arrowtypec #3
\arrowtyped #4
\width #5
\height #6
}
\def\setpos(#1,#2){\xpos=#1 \ypos#2}

\def\settriparms[#1`#2`#3;#4]{\settripairparms[#1`#2`#3`1`1;#4]}%
 
\def\settripairparms[#1`#2`#3`#4`#5;#6]{%
\arrowtypea #1
\arrowtypeb #2
\arrowtypec #3
\arrowtyped #4
\arrowtypee #5
\width #6
\height #6
}
 
\def\resetparms{\settripairparms[1`1`1`1`1;500]\width 500}%default values%
 
\resetparms
 
\def\mvector(#1,#2)#3{%%
\put(0,0){\vector(#1,#2){#3}}%
\put(0,0){\vector(#1,#2){26}}%
}
\def\evector(#1,#2)#3{{%%
\arrowlength #3
\put(0,0){\vector(#1,#2){\arrowlength}}%
\advance \arrowlength by-30
\put(0,0){\vector(#1,#2){\arrowlength}}%
}}
 
\def\horsize#1#2{%
\settowidth{\tempdimen}{$#2$}%
#1=\tempdimen
\divide #1 by\unitlength
}
 
\def\vertsize#1#2{%
\settoheight{\tempdimen}{$#2$}%
#1=\tempdimen
\settodepth{\tempdimen}{$#2$}%
\advance #1 by\tempdimen
\divide #1 by\unitlength
}
 
\def\putvector(#1,#2)(#3,#4)#5#6{{%
\ifnum3<\arrowtype
\putdashvector(#1,#2)(#3,#4)#5\arrowtype
\else
\ifnum\arrowtype<-3
\putdashvector(#1,#2)(#3,#4)#5\arrowtype
\else
\xpos=#1
\ypos=#2
\run=#3
\rise=#4
\arrowlength=#5
\ifnum \arrowtype<0
    \ifnum \run=0
        \advance \ypos by-\arrowlength
    \else
        \tempcounta \arrowlength
        \multiply \tempcounta by\rise
        \divide \tempcounta by\run
        \ifnum\run>0
            \advance \xpos by\arrowlength
            \advance \ypos by\tempcounta
        \else
            \advance \xpos by-\arrowlength
            \advance \ypos by-\tempcounta
        \fi
    \fi
    \multiply \arrowtype by-1
    \multiply \rise by-1
    \multiply \run by-1
\fi
\ifcase \arrowtype
\or \put(\xpos,\ypos){\vector(\run,\rise){\arrowlength}}%
\or \put(\xpos,\ypos){\mvector(\run,\rise)\arrowlength}%
\or \put(\xpos,\ypos){\evector(\run,\rise){\arrowlength}}%
\fi\fi\fi
}}
 
\def\putsplitvector(#1,#2)#3#4{%%
\xpos #1
\ypos #2
\arrowtype #4
\halflength #3
\arrowlength #3
\gap 140
\advance \halflength by-\gap
\divide \halflength by2
\ifnum\arrowtype>0
   \ifcase \arrowtype
   \or \put(\xpos,\ypos){\line(0,-1){\halflength}}%
       \advance\ypos by-\halflength
       \advance\ypos by-\gap
       \put(\xpos,\ypos){\vector(0,-1){\halflength}}%
   \or \put(\xpos,\ypos){\line(0,-1)\halflength}%
       \put(\xpos,\ypos){\vector(0,-1)3}%
       \advance\ypos by-\halflength
       \advance\ypos by-\gap
       \put(\xpos,\ypos){\vector(0,-1){\halflength}}%
   \or \put(\xpos,\ypos){\line(0,-1)\halflength}%
       \advance\ypos by-\halflength
       \advance\ypos by-\gap
       \put(\xpos,\ypos){\evector(0,-1){\halflength}}%
   \fi
\else \arrowtype=-\arrowtype
   \ifcase\arrowtype
   \or \advance \ypos by-\arrowlength
       \put(\xpos,\ypos){\line(0,1){\halflength}}%
       \advance\ypos by\halflength
       \advance\ypos by\gap
       \put(\xpos,\ypos){\vector(0,1){\halflength}}%
   \or \advance \ypos by-\arrowlength
       \put(\xpos,\ypos){\line(0,1)\halflength}%
       \put(\xpos,\ypos){\vector(0,1)3}%
       \advance\ypos by\halflength
       \advance\ypos by\gap
       \put(\xpos,\ypos){\vector(0,1){\halflength}}%
   \or \advance \ypos by-\arrowlength
       \put(\xpos,\ypos){\line(0,1)\halflength}%
       \advance\ypos by\halflength
       \advance\ypos by\gap
       \put(\xpos,\ypos){\evector(0,1){\halflength}}%
   \fi
\fi
}
 
\def\putmorphism(#1)(#2,#3)[#4`#5`#6]#7#8#9{{%
\run #2
\rise #3
\ifnum\rise=0
  \puthmorphism(#1)[#4`#5`#6]{#7}{#8}#9%
\else\ifnum\run=0
  \putvmorphism(#1)[#4`#5`#6]{#7}{#8}#9%
\else
\setpos(#1)%
\arrowlength #7
\arrowtype #8
\ifnum\run=0
\else\ifnum\rise=0
\else
\ifnum\run>0
    \coefa=1
\else
   \coefa=-1
\fi
\ifnum\arrowtype>0
   \coefb=0
   \coefc=-1
\else
   \coefb=\coefa
   \coefc=1
   \arrowtype=-\arrowtype
\fi
\width=2
\multiply \width by\run
\divide \width by\rise
\ifnum \width<0  \width=-\width\fi
\advance\width by60
\if l#9 \width=-\width\fi
\putbox(\xpos,\ypos){#4}%            %node 1
{\multiply \coefa by\arrowlength%      %node 2
\advance\xpos by\coefa
\multiply \coefa by\rise
\divide \coefa by\run
\advance \ypos by\coefa
\putbox(\xpos,\ypos){#5} }%
{\multiply \coefa by\arrowlength%      %label
\divide \coefa by2
\advance \xpos by\coefa
\advance \xpos by\width
\multiply \coefa by\rise
\divide \coefa by\run
\advance \ypos by\coefa
\if l#9%
   \putrbox(\xpos,\ypos){#6}%
\else\if r#9%
   \putlbox(\xpos,\ypos){#6}%
\fi\fi }%
{\multiply \rise by-\coefc%             %arrow
\multiply \run by-\coefc
\multiply \coefb by\arrowlength
\advance \xpos by\coefb
\multiply \coefb by\rise
\divide \coefb by\run
\advance \ypos by\coefb
\multiply \coefc by70
\advance \ypos by\coefc
\multiply \coefc by\run
\divide \coefc by\rise
\advance \xpos by\coefc
\multiply \coefa by140
\multiply \coefa by\run
\divide \coefa by\rise
\advance \arrowlength by\coefa
\ifcase\arrowtype
\or \put(\xpos,\ypos){\vector(\run,\rise){\arrowlength}}%
\or \put(\xpos,\ypos){\mvector(\run,\rise){\arrowlength}}%
\or \put(\xpos,\ypos){\evector(\run,\rise){\arrowlength}}%
\fi}\fi\fi\fi\fi}}

\newcount\numbdashes \newcount\lengthdash \newcount\increment
 
\def\howmanydashes{% Actually returns both number and length
\numbdashes=\arrowlength \lengthdash=40
\divide\numbdashes by \lengthdash
\lengthdash=\arrowlength
\divide\lengthdash by \numbdashes
\increment=\lengthdash
\multiply\lengthdash by 3
\divide\lengthdash by 5
}
 
\def\putdashvector(#1)(#2,#3)#4#5{%
\ifnum#3=0 \putdashhvector(#1){#4}#5
\else
\ifnum#2=0
\putdashvvector(#1){#4}#5\fi\fi}
 
\def\putdashhvector(#1,#2)#3#4{{%
\arrowlength=#3 \howmanydashes
\multiput(#1,#2)(\increment,0){\numbdashes}%
{\vrule height .4pt width \lengthdash\unitlength}
\arrowtype=#4 \xpos=#1
\ifnum\arrowtype<0 \advance\arrowtype by 7 \fi
\ifcase\arrowtype
\or \advance\xpos by 10
    \put(\xpos,#2){\vector(-1,0){\lengthdash}}
    \advance\xpos by 40
    \put(\xpos,#2){\vector(-1,0){\lengthdash}}
\or \advance \xpos by 10
    \put(\xpos,#2){\vector(-1,0){\lengthdash}}
    \advance\xpos by  \arrowlength
    \advance\xpos by  -50
    \put(\xpos,#2){\vector(-1,0){\lengthdash}}
\or \advance\xpos by 10
    \put(\xpos,#2){\vector(-1,0){\lengthdash}}
\or \advance\xpos by \arrowlength
    \advance\xpos by -\lengthdash
    \put(\xpos,#2){\vector(1,0){\lengthdash}}
\or {\advance\xpos by 10
    \put(\xpos,#2){\vector(1,0){\lengthdash}}}
    \advance\xpos by \arrowlength
    \advance\xpos by -\lengthdash
    \put(\xpos,#2){\vector(1,0){\lengthdash}}
\or \advance\xpos by \arrowlength
    \advance\xpos by -\lengthdash
    \put(\xpos,#2){\vector(1,0){\lengthdash}}
    \advance\xpos by -40
    \put(\xpos,#2){\vector(1,0){\lengthdash}}
   \fi
}}
 
\def\putdashvvector(#1,#2)#3#4{{%
\arrowlength=#3 \howmanydashes
\ypos=#2 \advance\ypos by -\arrowlength
\multiput(#1,#2)(0,\increment){\numbdashes}%
    {\vrule width .4pt height \lengthdash\unitlength}
\arrowtype=#4 \ypos=#2
\ifnum\arrowtype<0 \advance\arrowtype by 7 \fi
\ifcase\arrowtype
\or \advance\ypos by \arrowlength \advance\ypos by -40
    \put(#1,\ypos){\vector(0,1){\lengthdash}}
    \advance\ypos by -40
    \put(#1,\ypos){\vector(0,1){\lengthdash}}
\or \advance\ypos by 10
    \put(#1,\ypos){\vector(0,1){\lengthdash}}
    \advance\ypos by \arrowlength \advance\ypos by -40
    \put(#1,\ypos){\vector(0,1){\lengthdash}}
\or \advance\ypos by \arrowlength \advance\ypos by -40
    \put(#1,\ypos){\vector(0,1){\lengthdash}}
\or \advance\ypos by 10
    \put(#1,\ypos){\vector(0,-1){\lengthdash}}
\or \advance\ypos by 10
    \put(#1,\ypos){\vector(0,-1){\lengthdash}}
    \advance\ypos by \arrowlength \advance\ypos by -40
    \put(#1,\ypos){\vector(0,-1){\lengthdash}}
\or \advance\ypos by 10
    \put(#1,\ypos){\vector(0,-1){\lengthdash}}
    \advance\ypos by 40
    \put(#1,\ypos){\vector(0,-1){\lengthdash}}
\fi
}}
 
\def\puthmorphism(#1,#2)[#3`#4`#5]#6#7#8{{%
\xpos #1
\ypos #2
\width #6
\arrowlength #6
\arrowtype=#7
\putbox(\xpos,\ypos){#3\vphantom{#4}}%
{\advance \xpos by\arrowlength
\putbox(\xpos,\ypos){\vphantom{#3}#4}}%
\horsize{\tempcounta}{#3}%
\horsize{\tempcountb}{#4}%
\divide \tempcounta by2
\divide \tempcountb by2
\advance \tempcounta by30
\advance \tempcountb by30
\advance \xpos by\tempcounta
\advance \arrowlength by-\tempcounta
\advance \arrowlength by-\tempcountb
\putvector(\xpos,\ypos)(1,0)\arrowlength\arrowtype
\divide \arrowlength by2
\advance \xpos by\arrowlength
\vertsize{\tempcounta}{#5}%
\divide\tempcounta by2
\advance \tempcounta by20
\if a#8 %
   \advance \ypos by\tempcounta
   \putbox(\xpos,\ypos){#5}%
\else
   \advance \ypos by-\tempcounta
   \putbox(\xpos,\ypos){#5}%
\fi}}
 
\def\putvmorphism(#1,#2)[#3`#4`#5]#6#7#8{{%
\xpos #1
\ypos #2
\arrowlength #6
\arrowtype #7
\settowidth{\xlen}{$#5$}%
\putbox(\xpos,\ypos){#3}%
{\advance \ypos by-\arrowlength
\putbox(\xpos,\ypos){#4}}%
{\advance\arrowlength by-140
\advance \ypos by-70
\ifdim\xlen>0pt
   \if m#8%
      \putsplitvector(\xpos,\ypos)\arrowlength\arrowtype
   \else
   \putvector(\xpos,\ypos)(0,-1)\arrowlength\arrowtype
   \fi
\else
   \putvector(\xpos,\ypos)(0,-1)\arrowlength\arrowtype
\fi}%
\ifdim\xlen>0pt
   \divide \arrowlength by2
   \advance\ypos by-\arrowlength
   \if l#8%
      \advance \xpos by-40
      \putrbox(\xpos,\ypos){#5}%
   \else\if r#8%
      \advance \xpos by40
      \putlbox(\xpos,\ypos){#5}%
   \else
      \putbox(\xpos,\ypos){#5}%
   \fi\fi
\fi
}}
 
\def\putsquarep<#1>(#2)[#3;#4`#5`#6`#7]{{%
\setsqparms[#1]%
\setpos(#2)%
\settokens`#3`%
\puthmorphism(\xpos,\ypos)[\tokenc`\tokend`{#7}]{\width}{\arrowtyped}b%
\advance\ypos by \height
\puthmorphism(\xpos,\ypos)[\tokena`\tokenb`{#4}]{\width}{\arrowtypea}a%
\putvmorphism(\xpos,\ypos)[``{#5}]{\height}{\arrowtypeb}l%
\advance\xpos by \width
\putvmorphism(\xpos,\ypos)[``{#6}]{\height}{\arrowtypec}r%
}}
 
\def\putsquare{\@ifnextchar <{\putsquarep}{\putsquarep%
   <\arrowtypea`\arrowtypeb`\arrowtypec`\arrowtyped;\width`\height>}}
\def\square{\@ifnextchar< {\squarep}{\squarep
   <\arrowtypea`\arrowtypeb`\arrowtypec`\arrowtyped;\width`\height>}}
\def\squarep<#1>[#2`#3`#4`#5;#6`#7`#8`#9]{{%       %     #2------>#3
\setsqparms[#1]%                                   %      |       |
\diagram%                                          %      |       |
\putsquarep<\arrowtypea`\arrowtypeb`\arrowtypec`%  %    #7|       |#8
\arrowtyped;\width`\height>%                       %      |       |
(0,0)[#2`#3`#4`{#5};#6`#7`#8`{#9}]%                %      |       |
\enddiagram%                                       %      v       v
}}                                                 %     #4------>#5
\def\putptrianglep<#1>(#2,#3)[#4`#5`#6;#7`#8`#9]{{%
\settriparms[#1]%
\xpos=#2 \ypos=#3
\advance\ypos by \height
\puthmorphism(\xpos,\ypos)[#4`#5`{#7}]{\height}{\arrowtypea}a%
\putvmorphism(\xpos,\ypos)[`#6`{#8}]{\height}{\arrowtypeb}l%
\advance\xpos by\height
\putmorphism(\xpos,\ypos)(-1,-1)[``{#9}]{\height}{\arrowtypec}r%
}}
 
\def\putptriangle{\@ifnextchar <{\putptrianglep}{\putptrianglep
   <\arrowtypea`\arrowtypeb`\arrowtypec;\height>}}
\def\ptriangle{\@ifnextchar <{\ptrianglep}{\ptrianglep
   <\arrowtypea`\arrowtypeb`\arrowtypec;\height>}}
\def\ptrianglep<#1>[#2`#3`#4;#5`#6`#7]{{%%    %      #2----->#3
\settriparms[#1]%                             %      |      /
\diagram%                                     %      |     /
\putptrianglep<\arrowtypea`\arrowtypeb`%      %    #6|    /#7
\arrowtypec;\height>%                         %      |   /
(0,0)[#2`#3`#4;#5`#6`{#7}]%                   %      |  /
\enddiagram%%                                 %      v v
}}                                            %      #4
 
\def\putqtrianglep<#1>(#2,#3)[#4`#5`#6;#7`#8`#9]{{%
\settriparms[#1]%
\xpos=#2 \ypos=#3
\advance\ypos by\height
\puthmorphism(\xpos,\ypos)[#4`#5`{#7}]{\height}{\arrowtypea}a%
\putmorphism(\xpos,\ypos)(1,-1)[``{#8}]{\height}{\arrowtypeb}l%
\advance\xpos by\height
\putvmorphism(\xpos,\ypos)[`#6`{#9}]{\height}{\arrowtypec}r%
}}
 
\def\putqtriangle{\@ifnextchar <{\putqtrianglep}{\putqtrianglep
   <\arrowtypea`\arrowtypeb`\arrowtypec;\height>}}
\def\qtriangle{\@ifnextchar <{\qtrianglep}{\qtrianglep
   <\arrowtypea`\arrowtypeb`\arrowtypec;\height>}}
\def\qtrianglep<#1>[#2`#3`#4;#5`#6`#7]{{%%    %        #2----->#3
\settriparms[#1]%                             %         \      |
\width=\height                                %          \     |
\diagram%                                     %         #6\    |#7
\putqtrianglep<\arrowtypea`\arrowtypeb`%      %            \   |
\arrowtypec;\height>%                         %             \  |
(0,0)[#2`#3`#4;#5`#6`{#7}]%                   %              v v
\enddiagram%%                                 %               #4
}}
 
\def\putdtrianglep<#1>(#2,#3)[#4`#5`#6;#7`#8`#9]{{%
\settriparms[#1]%
\xpos=#2 \ypos=#3
\puthmorphism(\xpos,\ypos)[#5`#6`{#9}]{\height}{\arrowtypec}b%
\advance\xpos by \height \advance\ypos by\height
\putmorphism(\xpos,\ypos)(-1,-1)[``{#7}]{\height}{\arrowtypea}l%
\putvmorphism(\xpos,\ypos)[#4``{#8}]{\height}{\arrowtypeb}r%
}}
 
\def\putdtriangle{\@ifnextchar <{\putdtrianglep}{\putdtrianglep
   <\arrowtypea`\arrowtypeb`\arrowtypec;\height>}}
\def\dtriangle{\@ifnextchar <{\dtrianglep}{\dtrianglep
   <\arrowtypea`\arrowtypeb`\arrowtypec;\height>}}
\def\dtrianglep<#1>[#2`#3`#4;#5`#6`#7]{{%%    %                  / |
\settriparms[#1]%                             %                 /  |
\width=\height                                %              #5/   |#6
\diagram%                                     %               /    |
\putdtrianglep<\arrowtypea`\arrowtypeb`%      %              /     |
\arrowtypec;\height>%                         %             v      v
(0,0)[#2`#3`#4;#5`#6`{#7}]%                   %            #3----->#4
\enddiagram%%                                 %                #7
}}
 
\def\putbtrianglep<#1>(#2,#3)[#4`#5`#6;#7`#8`#9]{{%
\settriparms[#1]%
\xpos=#2 \ypos=#3
\puthmorphism(\xpos,\ypos)[#5`#6`{#9}]{\height}{\arrowtypec}b%
\advance\ypos by\height
\putmorphism(\xpos,\ypos)(1,-1)[``{#8}]{\height}{\arrowtypeb}r%
\putvmorphism(\xpos,\ypos)[#4``{#7}]{\height}{\arrowtypea}l%
}}
 
\def\putbtriangle{\@ifnextchar <{\putbtrianglep}{\putbtrianglep
   <\arrowtypea`\arrowtypeb`\arrowtypec;\height>}}
\def\btriangle{\@ifnextchar <{\btrianglep}{\btrianglep
   <\arrowtypea`\arrowtypeb`\arrowtypec;\height>}}
\def\btrianglep<#1>[#2`#3`#4;#5`#6`#7]{{%%   %              | \
\settriparms[#1]%                            %              |  \
\width=\height                               %            #5|   \#6
\diagram%                                    %              |    \
\putbtrianglep<\arrowtypea`\arrowtypeb`%     %              |     \
\arrowtypec;\height>%                        %              v      v
(0,0)[#2`#3`#4;#5`#6`{#7}]%                  %              #3----->#4
\enddiagram%%                                %                 #7
}}
 
\def\putAtrianglep<#1>(#2,#3)[#4`#5`#6;#7`#8`#9]{{%
\settriparms[#1]%
\xpos=#2 \ypos=#3
{\multiply \height by2
\puthmorphism(\xpos,\ypos)[#5`#6`{#9}]{\height}{\arrowtypec}b}%
\advance\xpos by\height \advance\ypos by\height
\putmorphism(\xpos,\ypos)(-1,-1)[#4``{#7}]{\height}{\arrowtypea}l%
\putmorphism(\xpos,\ypos)(1,-1)[``{#8}]{\height}{\arrowtypeb}r%
}}
 
\def\putAtriangle{\@ifnextchar <{\putAtrianglep}{\putAtrianglep
   <\arrowtypea`\arrowtypeb`\arrowtypec;\height>}}
\def\Atriangle{\@ifnextchar <{\Atrianglep}{\Atrianglep
   <\arrowtypea`\arrowtypeb`\arrowtypec;\height>}}
\def\Atrianglep<#1>[#2`#3`#4;#5`#6`#7]{{%%         %         /   \
\settriparms[#1]%                                  %        /     \
\width=\height                                     %     #5/       \#6
\diagram%                                          %      /         \
\putAtrianglep<\arrowtypea`\arrowtypeb`%           %     /           \
\arrowtypec;\height>%                              %    v             v
(0,0)[#2`#3`#4;#5`#6`{#7}]%                        %   #3------------>#4
\enddiagram%%                                      %          #7
}}
 
\def\putAtrianglepairp<#1>(#2)[#3;#4`#5`#6`#7`#8]{{%
\settripairparms[#1]%
\setpos(#2)%
\settokens`#3`%
\puthmorphism(\xpos,\ypos)[\tokenb`\tokenc`{#7}]{\height}{\arrowtyped}b%
\advance\xpos by\height
\puthmorphism(\xpos,\ypos)[\phantom{\tokenc}`\tokend`{#8}]%
{\height}{\arrowtypee}b%
\advance\ypos by\height
\putmorphism(\xpos,\ypos)(-1,-1)[\tokena``{#4}]{\height}{\arrowtypea}l%
\putvmorphism(\xpos,\ypos)[``{#5}]{\height}{\arrowtypeb}m%
\putmorphism(\xpos,\ypos)(1,-1)[``{#6}]{\height}{\arrowtypec}r%
}}
 
\def\putAtrianglepair{\@ifnextchar <{\putAtrianglepairp}{\putAtrianglepairp%
   <\arrowtypea`\arrowtypeb`\arrowtypec`\arrowtyped`\arrowtypee;\height>}}
\def\Atrianglepair{\@ifnextchar <{\Atrianglepairp}{\Atrianglepairp%
   <\arrowtypea`\arrowtypeb`\arrowtypec`\arrowtyped`\arrowtypee;\height>}}
 
\def\Atrianglepairp<#1>[#2;#3`#4`#5`#6`#7]{{%           %  #2a
\settripairparms[#1]%                         %           / | \
\settokens`#2`%                               %          /  |  \
\width=\height                                %       #3/  #4   \#5
\diagram%                                     %        /    |    \
\putAtrianglepairp                            %       /     |     \
<\arrowtypea`\arrowtypeb`\arrowtypec`%        %      v      v      v
\arrowtyped`\arrowtypee;\height>%             %     #2b---->#2c---->#2d
(0,0)[{#2};#3`#4`#5`#6`{#7}]%                 %         #6     #7
\enddiagram%%
}}
 
\def\putVtrianglep<#1>(#2,#3)[#4`#5`#6;#7`#8`#9]{{%
\settriparms[#1]%
\xpos=#2 \ypos=#3
\advance\ypos by\height
{\multiply\height by2
\puthmorphism(\xpos,\ypos)[#4`#5`{#7}]{\height}{\arrowtypea}a}%
\putmorphism(\xpos,\ypos)(1,-1)[`#6`{#8}]{\height}{\arrowtypeb}l%
\advance\xpos by\height
\advance\xpos by\height
\putmorphism(\xpos,\ypos)(-1,-1)[``{#9}]{\height}{\arrowtypec}r%
}}
 
\def\putVtriangle{\@ifnextchar <{\putVtrianglep}{\putVtrianglep
   <\arrowtypea`\arrowtypeb`\arrowtypec;\height>}}
\def\Vtriangle{\@ifnextchar <{\Vtrianglep}{\Vtrianglep
   <\arrowtypea`\arrowtypeb`\arrowtypec;\height>}}
\def\Vtrianglep<#1>[#2`#3`#4;#5`#6`#7]{{%%     %        #2------------->#3
\settriparms[#1]%                              %         \             /
\width=\height                                 %          \           /
\diagram%                                      %         #6\         /#7
\putVtrianglep<\arrowtypea`\arrowtypeb`%       %            \       /
\arrowtypec;\height>%                          %             \     /
(0,0)[#2`#3`#4;#5`#6`{#7}]%                    %              v   v
\enddiagram%%                                  %               #4
}}
 
\def\putVtrianglepairp<#1>(#2)[#3;#4`#5`#6`#7`#8]{{
\settripairparms[#1]%
\setpos(#2)%
\settokens`#3`%
\advance\ypos by\height
\putmorphism(\xpos,\ypos)(1,-1)[`\tokend`{#6}]{\height}{\arrowtypec}l%
\puthmorphism(\xpos,\ypos)[\tokena`\tokenb`{#4}]{\height}{\arrowtypea}a%
\advance\xpos by\height
\puthmorphism(\xpos,\ypos)[\phantom{\tokenb}`\tokenc`{#5}]%
{\height}{\arrowtypeb}a%
\putvmorphism(\xpos,\ypos)[``{#7}]{\height}{\arrowtyped}m%
\advance\xpos by\height
\putmorphism(\xpos,\ypos)(-1,-1)[``{#8}]{\height}{\arrowtypee}r%
}}
 
\def\putVtrianglepair{\@ifnextchar <{\putVtrianglepairp}{\putVtrianglepairp%
    <\arrowtypea`\arrowtypeb`\arrowtypec`\arrowtyped`\arrowtypee;\height>}}
\def\Vtrianglepair{\@ifnextchar <{\Vtrianglepairp}{\Vtrianglepairp%
    <\arrowtypea`\arrowtypeb`\arrowtypec`\arrowtyped`\arrowtypee;\height>}}
\def\Vtrianglepairp<#1>[#2;#3`#4`#5`#6`#7]{{%  %  #2a---->#2b---->#2c
\settripairparms[#1]%                          %   \      |      /
\settokens`#2`%                                %    \     |     /
\diagram%                                      %   #5\   #6    /#7
\putVtrianglepairp                             %      \   |   /
<\arrowtypea`\arrowtypeb`\arrowtypec`%         %       \  |  /
\arrowtyped`\arrowtypee;\height>%              %        v v v
(0,0)[{#2};#3`#4`#5`#6`{#7}]%                  %         #2d
\enddiagram%%
}}

\def\putCtrianglep<#1>(#2,#3)[#4`#5`#6;#7`#8`#9]{{%
\settriparms[#1]%
\xpos=#2 \ypos=#3
\advance\ypos by\height
\putmorphism(\xpos,\ypos)(1,-1)[``{#9}]{\height}{\arrowtypec}l%
\advance\xpos by\height
\advance\ypos by\height
\putmorphism(\xpos,\ypos)(-1,-1)[#4`#5`{#7}]{\height}{\arrowtypea}l%
{\multiply\height by 2
\putvmorphism(\xpos,\ypos)[`#6`{#8}]{\height}{\arrowtypeb}r}%
}}
 
\def\putCtriangle{\@ifnextchar <{\putCtrianglep}{\putCtrianglep
    <\arrowtypea`\arrowtypeb`\arrowtypec;\height>}}
\def\Ctriangle{\@ifnextchar <{\Ctrianglep}{\Ctrianglep
    <\arrowtypea`\arrowtypeb`\arrowtypec;\height>}}
\def\Ctrianglep<#1>[#2`#3`#4;#5`#6`#7]{{%%   %                / |
\settriparms[#1]%                            %             #5/  |
\width=\height                               %              /   |
\diagram%                                    %             v    |
\putCtrianglep<\arrowtypea`\arrowtypeb`%     %           #3     |#6
\arrowtypec;\height>%                        %             \    |
(0,0)[#2`#3`#4;#5`#6`{#7}]%                  %            #7\   |
\enddiagram%%                                %               \  |
}}                                           %                v v
\def\putDtrianglep<#1>(#2,#3)[#4`#5`#6;#7`#8`#9]{{%
\settriparms[#1]%
\xpos=#2 \ypos=#3
\advance\xpos by\height \advance\ypos by\height
\putmorphism(\xpos,\ypos)(-1,-1)[``{#9}]{\height}{\arrowtypec}r%
\advance\xpos by-\height \advance\ypos by\height
\putmorphism(\xpos,\ypos)(1,-1)[`#5`{#8}]{\height}{\arrowtypeb}r%
{\multiply\height by 2
\putvmorphism(\xpos,\ypos)[#4`#6`{#7}]{\height}{\arrowtypea}l}%
}}
 
\def\putDtriangle{\@ifnextchar <{\putDtrianglep}{\putDtrianglep
    <\arrowtypea`\arrowtypeb`\arrowtypec;\height>}}
\def\Dtriangle{\@ifnextchar <{\Dtrianglep}{\Dtrianglep
   <\arrowtypea`\arrowtypeb`\arrowtypec;\height>}}
\def\Dtrianglep<#1>[#2`#3`#4;#5`#6`#7]{{%%  %          | \
\settriparms[#1]%                           %          |  \#6
\width=\height                              %          |   \
\diagram%                                   %          |    v
\putDtrianglep<\arrowtypea`\arrowtypeb`%    %        #5|    #3
\arrowtypec;\height>%                       %          |    /
(0,0)[#2`#3`#4;#5`#6`{#7}]%                 %          |   /#7
\enddiagram%%                               %          |  /
}}                                          %          v v
\def\setrecparms[#1`#2]{\width=#1 \height=#2}%
\def\recursep<#1`#2>[#3;#4`#5`#6`#7`#8]{{\m@th
\width=#1 \height=#2
\settokens`#3`
\settowidth{\tempdimen}{$\tokena$}
\ifdim\tempdimen=0pt
  \savebox{\tempboxa}{\hbox{$\tokenb$}}%
  \savebox{\tempboxb}{\hbox{$\tokend$}}%
  \savebox{\tempboxc}{\hbox{$#6$}}%
\else
  \savebox{\tempboxa}{\hbox{$\hbox{$\tokena$}\times\hbox{$\tokenb$}$}}%
  \savebox{\tempboxb}{\hbox{$\hbox{$\tokena$}\times\hbox{$\tokend$}$}}%
  \savebox{\tempboxc}{\hbox{$\hbox{$\tokena$}\times\hbox{$#6$}$}}%
\fi
\ypos=\height
\divide\ypos by 2
\xpos=\ypos
\advance\xpos by \width
\bfig
\putCtrianglep<-1`1`1;\ypos>(0,0)[`\tokenc`;#5`#6`{#7}]%
\puthmorphism(\ypos,0)[\tokend`\usebox{\tempboxb}`{#8}]{\width}{-1}b%
\puthmorphism(\ypos,\height)[\tokenb`\usebox{\tempboxa}`{#4}]{\width}{-1}a%
\advance\ypos by \width
\putvmorphism(\ypos,\height)[``\usebox{\tempboxc}]{\height}1r%
\efig
}}
 
\def\recurse{\@ifnextchar <{\recursep}{\recursep<\width`\height>}}
 
\def\puttwohmorphisms(#1,#2)[#3`#4;#5`#6]#7#8#9{{%
\puthmorphism(#1,#2)[#3`#4`]{#7}0a
\ypos=#2
\advance\ypos by 20
\puthmorphism(#1,\ypos)[\phantom{#3}`\phantom{#4}`#5]{#7}{#8}a
\advance\ypos by -40
\puthmorphism(#1,\ypos)[\phantom{#3}`\phantom{#4}`#6]{#7}{#9}b
}}
 
\def\puttwovmorphisms(#1,#2)[#3`#4;#5`#6]#7#8#9{{%
\putvmorphism(#1,#2)[#3`#4`]{#7}0a
\xpos=#1
\advance\xpos by -20
\putvmorphism(\xpos,#2)[\phantom{#3}`\phantom{#4}`#5]{#7}{#8}l
\advance\xpos by 40
\putvmorphism(\xpos,#2)[\phantom{#3}`\phantom{#4}`#6]{#7}{#9}r
}}
 
\def\puthcoequalizer(#1)[#2`#3`#4;#5`#6`#7]#8#9{{%
\setpos(#1)%
\puttwohmorphisms(\xpos,\ypos)[#2`#3;#5`#6]{#8}11%
\advance\xpos by #8
\puthmorphism(\xpos,\ypos)[\phantom{#3}`#4`#7]{#8}1{#9}
}}
 
\def\putvcoequalizer(#1)[#2`#3`#4;#5`#6`#7]#8#9{{%
\setpos(#1)%
\puttwovmorphisms(\xpos,\ypos)[#2`#3;#5`#6]{#8}11%
\advance\ypos by -#8
\putvmorphism(\xpos,\ypos)[\phantom{#3}`#4`#7]{#8}1{#9}
}}
 
\def\putthreehmorphisms(#1)[#2`#3;#4`#5`#6]#7(#8)#9{{%
\setpos(#1) \settypes(#8)
\if a#9 %
     \vertsize{\tempcounta}{#5}%
     \vertsize{\tempcountb}{#6}%
     \ifnum \tempcounta<\tempcountb \tempcounta=\tempcountb \fi
\else
     \vertsize{\tempcounta}{#4}%
     \vertsize{\tempcountb}{#5}%
     \ifnum \tempcounta<\tempcountb \tempcounta=\tempcountb \fi
\fi
\advance \tempcounta by 60
\puthmorphism(\xpos,\ypos)[#2`#3`#5]{#7}{\arrowtypeb}{#9}
\advance\ypos by \tempcounta
\puthmorphism(\xpos,\ypos)[\phantom{#2}`\phantom{#3}`#4]{#7}{\arrowtypea}{#9}
\advance\ypos by -\tempcounta \advance\ypos by -\tempcounta
\puthmorphism(\xpos,\ypos)[\phantom{#2}`\phantom{#3}`#6]{#7}{\arrowtypec}{#9}
}}
 
\def\setarrowtoks[#1`#2`#3`#4`#5`#6]{%
\def\toka{#1}
\def\tokb{#2}
\def\tokc{#3}
\def\tokd{#4}
\def\toke{#5}
\def\tokf{#6}
}
\def\hex{\@ifnextchar <{\hexp}{\hexp<1000`400>}}
\def\hexp<#1`#2>[#3`#4`#5`#6`#7`#8;#9]{%
\setarrowtoks[#9]
\yext=#2 \advance \yext by #2
\xext=#1 \advance\xext by \yext
\bfig
\putCtriangle<-1`0`1;#2>(0,0)[`#5`;\tokb``\tokd]
\xext=#1 \yext=#2 \advance \yext by #2
\putsquare<1`0`0`1;\xext`\yext>(#2,0)[#3`#4`#7`#8;\toka```\tokf]
\advance \xext by #2
\putDtriangle<0`1`-1;#2>(\xext,0)[`#6`;`\tokc`\toke]
\efig
}
\makeatother

\newcommand{\PCPcontraction}{%
  \mathbin{%
    \hbox{%
      \!\vrule height 0.4pt width 5.0pt%
      }%
   \hbox{\vrule height 6.0pt width 0.4pt}\;
    }%
}%

\def\PCPqed{\hspace*{\fill}\ensuremath{\Box}\\}

\newcommand{\PCPwedge}{}
\newcommand{\PCPrestrict}[1]{%
        \raisebox{0ex}{\ensuremath{\restriction}}
        \hspace*{-0.52ex}\rule[-1.4ex]{0.09ex}{2.5ex}\,
        \raisebox{-1.4ex}{\ensuremath{\scriptscriptstyle #1}}}

\begin{document}

\begin{frontmatter}
  \title{  
%    \begin{flushright}
%      \small Freiburg THEP-??/??
%    \end{flushright}
%    \vspace{3ex}
    Geometry of Hamiltonean $n$-vectorfields in Multisymplectic Field Theory
    }
  
  \author{Cornelius Pauf\/ler and Hartmann R\"omer}
  \address{Fakult\"at f\"ur Physik\\
    Albert-Ludwigs-Universit\"at Freiburg im Breisgau\\
    Hermann-Herder-Stra\ss e 3\\
    D 79104 Freiburg i. Br.\\
    Germany\\
    e-mail: paufler@physik.uni-freiburg.de, roemer@physik.uni-freiburg.de
    }
  \date{\today}
  \begin{abstract}
    Multisymplectic geometry --- which originates from the well known
    De Donder--Weyl theory --- is a natural framework for the study of
    classical field theories. Recently, two algebraic structures have
    been put forward to encode a given theory algebraically. Those
    structures are formulated on finite dimensional spaces, which
    seems to be surprising at first.\\
    In this article, we investigate the correspondence of Hamiltonean
    functions and certain antisymmetric tensor products of
    vectorfields. 
    The latter turn out to be the proper generalisation of the
    Hamiltonean vectorfields of classical mechanics. Thus we clarify
    the algebraic description of solutions of the field equations.\\
  \end{abstract}
\end{frontmatter}
%%%%%%%%%%%%%%%%%%%%%%%%%%%%%%%%%%%%%%%%%%%%%%%%%%%%%%%%%%%%%%%%%%%%%%%%%%%%%
\section{Introduction}

It has long been known that the appropriate language for classical field
theories is the formalism of jet bundles. Within this framework, the
Langrangean variational principle can be formulated and the
Euler-Lagrange equations can be derived. 
Furthermore, the theorem of Emmy Noether (\cite{Noe18b}) which relates
symmetries of 
the Lagrange density and conserved quantities can be given a
geometrical interpretation.

In this article we consider first order field theories, i.e. theories
which are defined by a Lagrange density that depends on the fields and
their first derivatives only. In this case the field equations are
second order partial differential equations. These equations can be
transformed into a Hamiltonean system on an
infinite dimensional space --- this is the canonical Hamiltonean
formalism on the space of initial data. One has to distinguish a time
direction in order to define a conjugate momentum for every field
coordinate. This results in breaking Lorentz covariance.\\ 
Alternatively, there is a
framework that can be formulated on finite dimensional geometries (for
 a detailed review, we refer to \cite{GoIsMa98}). Moreover,
space and time directions are treated in a covariant way.
This approach is known under the name De Donder--Weyl
(DW) formalism or covariant Hamiltonean theory.  The article at hand
will stay within this framework.\\
In contrast to classical mechanics, it introduces more
than one conjugated momentum variable for each degree of freedom. 
Using a covariant generalisation of the Legendre
transformation of classical mechanics,
one can perform the transition from the second order Euler-Lagrange
equations to the first order DW equations. The latter are
formulated for sections of what is called multisymplectic phase space,
i.e. smooth maps from the base manifold into that space. Keeping in
mind that trajectories in classical mechanics are maps from the time
axis to phase space, the treatment in the DW formalism is a
generalisation to more than one evolution parameter.

Only recently two algebraic structures have been proposed that encode
the up to now geometrical picture of (partial) differential equations
for sections. While Forger and R\"omer (\cite{ForgerRoemer:2000}) work
on the extended multisymplectic phase space $\mathcal P$ that
generalises the doubly extended phases space of time dependent
symplectic mechanics, Kanatchikov (\cite{Kanatchikov:1997}) uses a
space that has one dimension less than $\mathcal P$ and can be
interpreted as the parameter space of hypersurfaces of constant De
Donder-Weyl Hamiltoneans. This space will be denoted $\tilde{\mathcal
  P}$ for the rest of this article.\\
Note that both $\mathcal P$ and $\tilde{\mathcal P}$ are
multisymplectic manifolds in the sense of Martin (\cite{Martin:1988a},
in which there is a generalised Darboux theorem)
only for very special bundles $\mathcal E$. Rather, we will use the term in
the more general sense of a manifold with a closed, non-degenerate
form (\cite{GoIsMa98,CantrijnIbortDeLeon:1999}).\\ 
The main advantage as compared to ordinary field theoretical Poisson
structures is that the underlying spaces $\mathcal P$ and
$\tilde{\mathcal P}$ both are finite dimensional. The
price one has to pay for this is that there is more than one
conjugated momentum associated with each coordinate degree of
freedom. Up to now, this has been an obstacle to the application of
the standard quantisation programme.

It remains to understand in which sense the algebraic structures 
desribe the solutions of the field equations, i.e. the states of the
system under consideration.

The idea is that in the case of mechanics there is a
correspondence between vectorfields and and curves in phase
space. The former can be viewed as derivations on the algebra of
smooth functions on the phase space, and can be described by a
functions that act via the Poisson bracket if the vectorfields are
Hamiltonean. 
In multisymplectic geometry, on the other hand, 
curves are replaced by sections of some bundle which
consequently are higher dimensional. Therefore, they are
described by a set of tangent vectors at every point 
which span a distribution on the extended multisymplectic phase space,
i.e. that specify at every point some subspace of the tangent
space. Furthermore, if the distribution is of constant rank (i.e. if
the sections do not have kinks), one can pick (in a smooth way) a basis of the
specified subspace in the tangent bundle at every point and combine
the basis vectors using the antisymmetric tensor product of vectors to
obtain a multivectorfield, see figure 1. This multivectorfield is unique up to
multiplication by a function and of constant tensor grade.
\begin{figure}[h]
  \begin{center}
    \includegraphics[width=6cm]{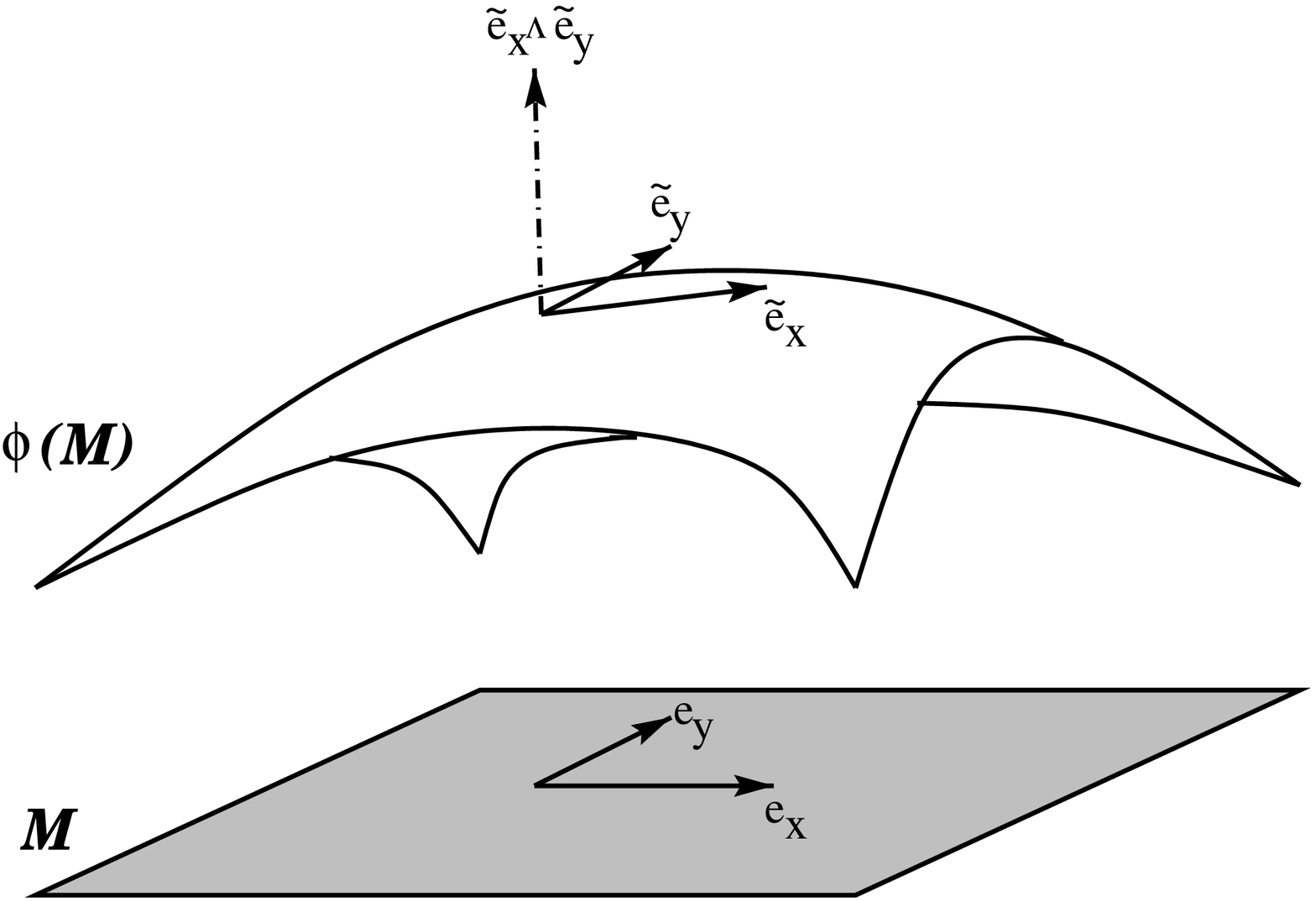}
  \end{center}
  \caption{The lifted vectors $\tilde e_x$ and $\tilde e_y$ span the
    tangent space at a point of the section $\Phi$. The two vectors can be
    combined to give a $2$-vector $\tilde e_x\wedge\tilde e_y$
    that describes this tangent space.}
\end{figure}

Kanatchikov was the first to note that the fundamental relationship of
symplectic geometry between Hamiltonean vectorfields $X_f$ and
functions $f$ given by
\begin{equation}\label{Xw-df}
  X_f\PCPcontraction \omega=df,
\end{equation}
where $\omega$ denotes the symplectic $2$-form, can be generalised to
cover the multisymplectic case, in which $\omega$ is the
multisymplectic form, a closed non-degenerate $(n+1)$-form
($n$ being the dimension of space-time), $f$ is an $r$-form and
$X_f$ has to be a multivectorfield of tensor grade
$(n-r)$. Consequently, if $f$ is a function then $X_f$ has to be an
$n$-vectorfield, i.e. a multivectorfield of tensor grade $n$. This is
a good candidate to describe distributions that yield sections of the
fibre bundle.
The link between Hamiltonean $n$-vectorfields and solutions of the 
field equations has already been indicated by Kanatchikov in
\cite{Kanatchikov:1997}. Moreover, the sense in which
multivectorfields are related to distributions seems to folklore and
is written out explicitely in the work by
Echeverr\'{\i}a-Enr\'{\i}quez et al. (\cite{EeMlRr}). However, both
use the smaller multisymplectic phase space $\tilde{\mathcal P}$ which
requires the choice of a connection, \cite{Pa00}. 
Moreover, we will show in
\ref{no-go} that for typical cases in field theory the generalisation
of (\ref{Xw-df}) to $\tilde{\mathcal P}$ does not admit the
interpretation of $X_f$ to define a distribution. Instead, one has to
go over to the extended multisymplectic phase space $\mathcal P$. This is
not in contradiction to the results established by
Echeverr\'{\i}a-Enr\'{\i}quez et al. since they consider an equation
different from (\ref{Xw-df}), namely
\begin{equation}
  X_f\PCPcontraction (\omega - df\wedge d^nx)=0,
\end{equation}
where $d^nx$ is a volume form on space-time (for non-trivial fibre
bundles, terms containing a connection appear in addition). 
Therefore, although their investigation proceeds along similar lines
as this article, the results cannot be taken over to the case of
multisymplectic geometry.

The structure of this article is as follows. Chapter \ref{geom} 
reviews the basic notions needed in this article. In
particular, the multisymplectic phase spaces $\mathcal P$ and 
$\tilde{\mathcal P}$ and the multisymplectic forms on them 
are defined and Hamiltonean forms and Hamiltonean
multivectorfields are introduced. 
Chapter \ref{n-vect} contains the main part of this article. We
will establish the link between solutions of the field equations and
multivectorfields associated to some appropriately chosen function in
three steps. Firstly, we show that a certain class of functions on
$\mathcal P$ admits Hamiltonean $n$-vectorfields that define
distributions. Secondly, we show that the leaves of those
distributions, should they exist, are solutions to the field equations
that correspond to the Hamiltonean function which has been chosen in
the first place. Thirdly, we investigate under which conditions the
distributions defined by the Hamiltonean $n$-vectorfields are
integrable. It will turn out that additional input is needed to answer
the latter question as there is a considerable freedom to choose a
Hamiltonean
$n$-vectorfield for a given Hamiltonean function.
This additional input is provided by a covariant version of the
Hamilton-Jacobi equation.
In the end, we will show that the construction cannot be take over to
$\tilde{\mathcal P}$.

%%%%%%%%%%%%%%%%%%%%%%%%%%%%%%%%%%%%%%%%%%%%%%%%%%%%%%%%%%%%%%%%%%%%%%%%%%%%%
\section{Multisymplectic geometry\label{geom}}

\subsection{De Donder-Weyl equations and multisymplectic phase spaces}

Usually, classical field theories are formulated as variational
problems for the fields $\varphi$ --  which are sections of
some fibre (vector) bundle $\mathcal E$ over an $n$-dimensional base manifold
(space-time) $\mathcal M$ -- and some Lagrange density $\mathcal L$.
The latter is a function of the fields and its first derivatives, and
one is looking for extremal points of the action functional
\begin{equation}
  \mathcal S(\varphi)=
  \int_{\mathcal M} \;d^nx\;\mathcal L(x,\varphi(x),\varphi'(x)).
\end{equation}
Mathematically, $\mathcal L$ is a function on the first jet bundle
$\mathfrak J^1\mathcal E$ to $\mathcal E$ (\cite{Sau89,GoIsMa98,KMS93}).
It is well known that the extremal points of this functional can be
found by solving the field equations -- the celebrated Euler-Lagrange
equations --
\begin{equation}\label{ELeq}
  \partial_\mu(\frac{\partial\mathcal
    L}{\partial_\mu\varphi^A}(\varphi(x))
  -\frac{\partial\mathcal  L}{\partial\varphi^A}(\varphi(x))=0.
\end{equation}
Here, as in all what follows, $\mu,\nu,\rho,\ldots=1,\ldots,n$ label
coordinates on $\mathcal M$, while $A,B,C,\ldots=1,\ldots,N$ stand for
those on the fibres of $\mathcal E$. 

If the Lagrange density fulfils some regularity condition, the
Euler-Lagrange equations can be seen to be equivalent to a set of
first order equations (cf. \cite{Ru66})
\begin{equation}\label{dDWHeq}
  \frac{\partial \mathcal H}{\partial \pi^\mu_A}(x,\varphi(x),\pi(x))
  =\partial_\mu\varphi^A(x),\quad
  \frac{\partial \mathcal H}{\partial \varphi^A}(x,\varphi(x),\pi(x))
  =-\partial_\mu\pi^\mu_A(x),
\end{equation}
for the DW Hamiltonean $\mathcal H$,
\begin{equation}\label{dDWHamiltonean}
  \mathcal H=\pi^\mu_A\partial_\mu\varphi^A-\mathcal L.
\end{equation}
In these equations, the polymomenta $\pi^\mu_A$ are defined as derivatives of
the Lagrange density by the field derivatives,
\begin{equation}\label{Polymomenta}
  \pi^\mu_A=\frac{\partial\mathcal L}{\partial\partial_\mu\varphi^A}.
\end{equation}
The regularity conditions to $\mathcal L$ ensure that these equations
can be used to express the field derivatives $\partial_\mu\varphi^A$
in terms of the fields $\varphi$ and the polymomenta.

So far we have used sections $\varphi(x)$ etc. to formulate the
equations of motion but it is useful to consider functions like the DW
Hamiltonean $\mathcal H$
without evaluating them on fields
$\varphi(x)$ etc.
To this end, let us introduce coordinates $v^A$ for fields, $v^A_\mu$
for 
their space-time derivatives and $p^\mu_A$ for the polymomenta
functions of (\ref{Polymomenta}). To
condense notation, we will write derivatives w.r.t. the field
$\varphi^A$ as $\partial_A$, while those w.r.t. the polymomenta
$\pi^\mu_A$ will be denoted by $\partial_\mu^A$. Together with an
additional coordinate $p$, the set of variables 
\begin{equation}\label{ExtPolPhSpCoord}
  (x^\mu,v^A,p^\mu_A,p)
\end{equation}
labels locally the extended multisymplectic phase space $\mathcal
P$. Derivatives by 
this extra coordinate, which itself can be interpreted as the De
Donder-Weyl energy
variable, will be denoted by $\partial$. Geometrically,
$\mathcal P$ is the (affine) dual of the first jet bundle $\mathfrak
J^1\mathcal E$, i.e. the space of fields and velocities. 
One can show that the choice of a local
chart of $\mathcal E$ induces coordinates on $\mathcal P$. The set of
coordinates 
\begin{equation}\label{PolPhSpCoord}
  (x^\mu,v^A,p^\mu_A)
\end{equation}
can be used to describe locally the multisymplectic phase space
$\tilde{\mathcal P}$. There 
is a canonical projection from $\mathcal P$ to $\tilde{\mathcal P}$ which projects
out the additional variable $p$. With the help of a volume form $\omega$ on
the base manifold $\mathcal M$ we find
\begin{equation}
  \tilde{\mathcal P}\stackrel{\omega}\cong (\mathfrak V\mathcal E)^\ast
  \otimes T\mathcal M,
  \quad
  \mathcal P\stackrel{\omega,\Gamma}\cong
  \tilde{\mathcal P}\oplus
  \mathcal R,
\end{equation}
where $\mathfrak V\mathcal E$ is the vertical tangent subbundle of
$\mathcal E$ and $\mathcal R$ denotes a trivial line bundle on
$\mathcal E$. For the latter isomorphism, a connection $\Gamma$ of
$\mathcal E$ is needed in addition (\cite{Pa00}). Note that the tensor
products are 
understood pointwise on $\mathcal E$.
 
At this point it is useful to examine the
special case if $\mathcal M$ happens to be the real axis $\mathbbm R$,
i.e. if there is only time and no space-like direction. Then,
$\mathcal E$ is trivial, say $\mathcal E=\mathbbm R\times \mathcal Q$,
and $\mathfrak J^1\mathcal E=\mathbbm R\times T\mathcal Q$. The
extended multisymplectic phase space $\mathcal P$ then becomes 
$\mathcal P=\mathbbm R^2\times T^\ast\mathcal Q$, which is the doubly
extended phase space of a time-dependent classical mechanical system with
configuration space $\mathcal Q$. $\tilde{\mathcal P}$ is in this case the
singly extended phase space. We will, however, suppress the word
single in order to keep the names short.

With these spaces introduced, 
equations (\ref{dDWHamiltonean}) and (\ref{Polymomenta}) can be
understood as a map  
\begin{equation}\label{Legendre}
  \mathbbm F\mathcal L:\mathfrak J^1\mathcal E\rightarrow \mathcal P,
  (x^\mu,v^A,v^A_\mu)\rightarrow 
  (x^\mu,v^A,\frac{\partial\mathcal L}{\partial v^A_\mu},
  -(v^A_\mu\frac{\partial\mathcal L}{\partial v^A_\mu}-\mathcal L)),
\end{equation}
which is known as Legendre transformation (the symbol $\mathbbm
F\mathcal L$ is chosen to express that it is a fibre derivation using
the Lagrange density). If the Lagrange density is
regular, this map defines a bijective map from 
$\mathfrak J^1\mathcal E$ to $\tilde{\mathcal P}$.

For convenience, the different spaces introduced so far will be displayed
in a diagram (the map $T$ will be needed below).
\begin{equation}
  \label{Diagramm}
  \begin{picture}(2100,1200)
    \putmorphism(1000,500)(0,-1)[\mathcal E`\mathcal M`{}]{500}{1}{l}
    \putmorphism(2000,1000)(-1,-1)[\frak J_1(\mathcal E)`
         \phantom{\mathcal M}`
         {}]{1000}{1}{r}
    \putmorphism(2000,1000)(-2,-1)[\phantom{\frak J^1(\mathcal E)}`
         \phantom{\mathcal M}`
         %{\scriptstyle \pi^1_0}
    ]{1000}{1}{r}
    \putmorphism(0,1000)(1,-1)[\tilde{\mathcal P}`
         \phantom{\mathcal M}`
         %{\scriptstyle \pi_{\scriptscriptstyle \mathcal E\mathcal P}}
    ]{1000}{1}{r}
    \putmorphism(1000,1000)(1,0)[\mathcal P=\mathfrak J_1^\ast(\mathcal E)`
         \phantom{\frak J_1(\mathcal E)}`
         {\scriptstyle \mathbbm F\mathcal L}
    ]{1000}{-1}{u}
    \putmorphism(0,1000)(1,0)[\phantom{\tilde{\mathcal P}}`\phantom{\mathcal P=\mathfrak J_1^\ast(\mathcal E)}`
         %{\scriptstyle \pi_{\scriptscriptstyle \mathcal P\Pi}}
    ]{1000}{-1}{r}
    \putmorphism(0,1000)(2,-1)[\phantom{\tilde{\mathcal P}}`\phantom{\mathcal E}`
         %{\scriptstyle \pi_{\scriptscriptstyle \mathcal M\mathcal P}}
    ]{1000}{1}{l}
    \putmorphism(1020,1000)(0,1)[\phantom{\mathcal P=\mathfrak J_1^\ast(\mathcal E)}`
         \phantom{\mathcal E}`\pi_{\mathcal E\mathcal P}
         %{\scriptstyle\pi_{\scriptscriptstyle \mathcal E\Pi}}
    ]{500}{1}{r}
    \putmorphism(980,1000)(0,1)[\phantom{\mathcal P=\mathfrak J_1^\ast(\mathcal E)}`
         \phantom{\mathcal E}`T
         %{\scriptstyle\pi_{\scriptscriptstyle \mathcal E\Pi}}
    ]{500}{-1}{l}
  \end{picture}
\end{equation}

%%%%%%%%%%%%%%%%%%%%%%%%%%%%%%%%%%%%%%%%%%%%%%%%%%%%%%%%%%%%%%%%%%%%%%%%%%%%%%
%\section{Algebraic structures\label{alg}}

\subsection{Multisymplectic forms}

It has long been known that there are generalisations of 
symplectic geometry to field theory. The crucial observation which
lead to the development of those generalisations was that in field
theory, solutions are sections (of some fibre bundle), while in
classical mechanics, solutions are curves. Hence, one can try to treat
the sections as higher dimensional analogues of curves, i.e. treat
the space-like coordinates of the fields under investigation as
additional evolution parameters, cf. (\ref{dDWHeq}). These efforts
culminated in the discovery of the multisymplectic form, an $n+1$-form which is to
replace the symplectic $2$-form.
The multisymplectic $n+1$-form is defined on the doubly extended
multisymplectic phase space $\mathcal P$. In coordinates, it is given by
\begin{equation}\label{FullPolyForm}
  \Omega_{(x,v,\vec p,p)}=dv^A\wedge dp^\mu_A\wedge d_\mu x-dp\wedge dx.
\end{equation}
Here, $\vec p$ is a shorthand notation for the polymomenta $p^\mu_A$.
We refer to the work of Gotay, 
Isenberg, Marsden et al. (\cite{GoIsMa98}) for a detailed
review. 
Note that $\Omega$ is an exact form.
Using $\Omega$, one defines pairs of 
Hamiltonean multivectorfields $X$, $X\in\Gamma(\Lambda^\ast T\mathcal P)$, 
and Hamiltonean forms $H$ by the
equation
\begin{equation}\label{DefHamVf}
  X\PCPcontraction\Omega = dH.
\end{equation}
From degree counting, it is immediate that $H$ can be a form of
maximal degree $(n-1)$. If $H$ is a homogeneous form of degree $|H|$,
then the corresponding Hamiltonean multivectorfield $X$ 
has to be an $(n-|H|)$-vectorfield. 
Observe that -- in contrast to classical mechanics --  
neither side is uniquely defined, although $\Omega$ is non degenerate
on vectorfields.

Because of the peculiar combination of field and polymomentum forms in
(\ref{FullPolyForm}) the dependence of a Hamiltonean form 
on the coordinates $p^\mu_A$ is
subject to strong restrictions. Unless $H$ is a function, it has to be
a polynomial of maximal degree $|H|$ in the polymomenta
(\cite{Kanatchikov:1997,Pa00}). 
There are additional restrictions to the specific form
of that polynomial dependence.

On the multisymplectic phase space $\tilde{\mathcal P}$, there is no such
canonical $(n+1)$-form, but on can separate the first summand of
(\ref{FullPolyForm}) and transport it to $\tilde{\mathcal P}$. The resulting
$(n+1)$-form is called vertical multisymplectic $(n+1)$-form. Its
coordinate expression is
\begin{equation}
  {\Omega_\Gamma}_{(x,v,\vec p,p)}=dv^A\wedge dp^\mu_A\wedge d_\mu x
  +{f_A}{\scriptstyle(x,v)}\,dv^A\wedge d^nx
  +{g_\mu^A}{\scriptstyle(x,v,\vec p)}\,dp^\mu_A\wedge d^nx.
\end{equation}
For the construction of $\Omega_\Gamma$, a connection of the fibre
bundle $\mathcal E$ has to be chosen. This choice creates the
last two terms in the above formula. Their precise expressions will not
be important for what follows (for them, we
refer to \cite{EeMlRr}). Using $\Gamma$ again, one can define a
vertical exterior derivative $d_\Gamma$ on $\tilde{\mathcal P}$, i.e. a map
with square zero that takes derivatives with respect to the $v^A$ and
$p^\mu_A$ variables only. Combining $\Omega_\Gamma$ and $d_\Gamma$,
one can ask for solutions $(X_H, H)$ of
\begin{equation}
  X_H\PCPcontraction \Omega_\Gamma=d_\Gamma H.
\end{equation}
In this case, $H$ is called Hamiltonean form on $\tilde{\mathcal P}$. Again,
the polymomentum dependence of $H$ is subject to restrictions unless
$H$ is a function.

%%%%%%%%%%%%%%%%%%%%%%%%%%%%%%%%%%%%%%%%%%%%%%%%%%%%%%%%%%%%%%%%%%%%%%%%%%%%%%
\section{Hamiltonean $n$-vectorfields\label{n-vect}}

\subsection{Decomposition of Hamiltonean $n$-vectorfields}

It is a well-known fact (\cite{KMS93}) that submanifolds can be
described by (integrable) distributions, i.e. the determination of
some subspace of the tangent bundle at every point of a manifold. In
the appendix, we show that such subspaces of dimension $n$ are in
correspondence to exactly the decomposable\footnote{
  There seems to be no standard terminology for the special elements
  in the $n$-fold antisymmetric tensor product of a vector space $V$
  that are of the form 
  \begin{equation}
    Z_1\wedge\cdots\wedge Z_n\in \Lambda^n(V), \quad Z_\mu\in V.
  \end{equation}
  In \cite{Greub:1978}, ch. 3, they are called {\em decomposable}, 
  while in \cite{DodsonPoston:1979}, section V.1.06, the word
  {\em simple} is used for them.
  }
$n$-vectorfields, i.e. such
vectorfields that can be written (locally) as the anti-symmetrised
tensor product of $n$ distinct vectorfields, cf. fig. 1. 
As explained in the appendix, $n$-dimensional subspaces of $T\mathcal P$ are
described by such $n$-vectorfields that can be written as the
$n$-fold antisymmetric tensor product of vectorfields. 
Therefore, we will examine for which
Hamiltonean $n$-vectorfields this property can be achieved.
\begin{prop}\label{prop:decomposition}
  Let $H\in \mathcal C^\infty(\mathcal P)$ be a function on the
  multisymplectic phase space. If $H$ is of the particular form
  \begin{equation}\label{goodH}
    H(x,v,\vec p,p)=-\mathcal H(x,v,\vec p)-p,
  \end{equation}
  where $\mathcal H$ is an arbitrary function not depending on $p$, 
  then there is a decomposable Hamiltonean vectorfield $X$ corresponding
  to $H$. 
%  The same is true if $H(x,v,\vec p,p)=-\mathcal H(x,v,\vec p)+p$ and 
%  if in addition there is a distinguished non vanishing vectorfield
%  on $\mathcal M$ (the time direction).
\end{prop}
{\bf Remark.} The condition on $H$ can be formulated without referring
to coordinates. 
As $\mathcal P$ is an affine bundle over $\tilde{\mathcal P}$ with a
trivial associated line bundle it
carries a fundamental vectorfield $\xi$, the derivation w.r.t. the
$p$-direction. The condition on $H$ is then 
%$\xi(H)>0$ and 
$\xi(H)=-1$%respectively
. It will become clear in the next section why we
distinguish the particular $p$-dependence. Note in particular that this
property does not depend on the coordinate system used.

{\bf Proof.} When expressed in coordinates,  
the condition for $X$ to be a Hamiltonean $n$-vectorfield to some
Hamiltonean $H\in\mathcal C^\infty(\mathcal P)$,
\begin{equation}\label{X}
  \begin{split}
    X\PCPcontraction\Omega&=dH,\textrm{ where }\\
    X&=
    {\scriptstyle\frac1{n!}}X^{\nu_1\cdots\nu_n}
    \partial_{\nu_1}\cdots\partial_{\nu_n}
    +
    {\scriptstyle\frac1{(n-1)!}}X^{A\nu_1\cdots\nu_{n-1}}
    \partial_A\PCPwedge\partial_{\nu_1}\cdots\partial_{\nu_{n-1}}\\
    &\quad
    +{\scriptstyle\frac1{(n-1)!}}{X^\sigma_A}^{\nu_1\cdots\nu_{n-1}}
    \partial_\sigma^A\PCPwedge\partial_{\nu_1}\cdots\partial_{\nu_{n-1}} 
    +{\scriptstyle\frac1{(n-1)!}}X^{\nu_1\cdots\nu_{n-1}}_0
    \partial\partial_{\nu_1}\cdots\partial_{\nu_{n-1}}
    \\
    &\quad
    +{\scriptstyle\frac1{(n-2)!}}{X^\sigma_A}^{B\nu_1\cdots\nu_{n-2}}
    \partial_\sigma^A\partial_B\PCPwedge\partial_{\nu_1}\cdots\partial_{\nu_{n-2}}
    +\textrm{terms of higher vertical order},
  \end{split}
\end{equation}
amounts to
\begin{equation}\label{HamVf-cond}
  \begin{split}
    \partial_AH&=    
    {\scriptstyle\frac{(-)^{n}}{(n-1)!} }
    {X^\mu_A}^{\nu_1\cdots\nu_{n-1}}
    \epsilon_{\mu\nu_1\cdots\nu_{n-1}}
    \\
    \partial_\mu^AH&=
    -{\scriptstyle\frac{(-)^n}{(n-1)!}}X^{A\nu_1\cdots\nu_{n-1}}
    \epsilon_{\mu\nu_1\cdots\nu_{n-1}}
    \\
    \partial_\mu H
    &=
    -{\scriptstyle\frac{1}{(n-2)!}}
    {X^\sigma_A}^{A\nu_1\cdots\nu_{n-2}}
    \epsilon_{\sigma\nu_1\cdots\nu_{n-2}\mu}
    -{\scriptstyle\frac{1}{(n-1)!}}
    X^{\nu_1\cdots\nu_{n-1}}_0\epsilon_{\nu_1\cdots\nu_{n-1}\mu}
    \\
    \partial H&={\scriptstyle\frac{(-)^{n+1}}{n!}}X^{\nu_1\cdots\nu_n}
    \epsilon_{\nu_1\cdots\nu_n}.
  \end{split}
\end{equation}
Now let $Z_\mu$ be a set of $n$ vectors,
\begin{equation}\label{defZmu}
  Z_\mu=(Z_\mu)^\nu\partial_\nu+(Z_\mu)^A\partial_A+(Z_\mu)^\nu_A\partial^A_\nu
  +(Z_\mu)_0\partial.
\end{equation}
The wedge product of all $Z_\mu$, $\mu=1,\ldots,n$ gives (in obvious
cases we will omit the symbol $\wedge$)
\begin{equation}\label{Y}
  \begin{split}
    Y&=Z_1\wedge \cdots\wedge Z_n\\
    &=
    (Y_1)^{\mu_1}\cdots(Y_n)^{\mu_n}\epsilon_{\mu_1\cdots\mu_n}
    \partial_{x^1}\PCPwedge\cdots\PCPwedge \partial_{x^n}
    \\&\quad
    +\sum_{\mu=1}^n(-)^{\mu+1}
    (Z_\mu)^A(Z_1)^{\nu_1}\cdots\widehat{(Z_\mu)^{\nu_\mu}}\cdots(Z_n)^{\nu_n}
    \partial_A\partial_{\nu_1}\cdots\widehat{\partial_{\nu_\mu}}\cdots\partial_{\nu_n}
    \\&\quad
    +\sum_{\mu=1}^n(-)^{\mu+1}
    (Z_\mu)^\sigma_A(Z_1)^{\nu_1}\cdots\widehat{(Z_\mu)^{\nu_\mu}}\cdots(Z_n)^{\nu_n}
    \partial_\sigma^A
    \partial_{\nu_1}\cdots\widehat{\partial_{\nu_\mu}}\cdots\partial_{\nu_n}
    \\&\quad
    +\sum_{\mu<\nu}
    (-)^{\mu+\nu}
    \big((Z_\mu)^A(Z_\nu)^\sigma_B-(Z_\nu)^A(Z_\mu)^\sigma_B)\big)
    \\&\quad\quad\quad\quad\quad\quad
    (Z_1)^{\rho_1}\cdots\widehat{(Z_\mu)^{\rho_\mu}}
    \cdots\widehat{(Z_\nu)^{\rho_\nu}}\cdots(Z_n)^{\rho_n}
    \partial_\sigma^B\partial_A
    \partial_{\rho_1}\cdots\widehat{\partial_{\rho_\mu}}
    \cdots\widehat{\partial_{\rho_\nu}}\cdots\partial_{\rho_n}
    \\
    &\quad+\textrm{terms of higher vertical order}
  \end{split}
\end{equation}
(In this calculation, a hat on top of a symbol means the omission of
that symbol.)

Comparing this to $X$, one finds in the first place
\begin{equation}\label{detZ}
  \begin{split}
    (Z_1)^{\mu_1}\cdots (Z_n)^{\mu_n}\epsilon_{\mu_1\cdots\mu_n}
    &=
    \frac1{n!}X^{\mu_1\cdots\mu_n}\epsilon_{\mu_1\cdots\mu_n}
    =(-)^{n+1}\partial H=(-1)^{n}
  \end{split}
\end{equation}
The $n$ vectors $Z_\mu$ of (\ref{defZmu}) define a linear map from
$T\mathcal M$ to $T\mathcal P$. Let us denote this map by $Z$. Using the
canonical projection $\pi_0^\star$ of $\mathcal P$ onto $\mathcal M$ we
obtain a map 
$T\pi_0^\star\circ Z$ from $T\mathcal M$ to itself. Equation
(\ref{detZ}) describes the determinant of this map in the coordinates
chosen. There is a straightforward solution, namely
\begin{equation}\label{Zmunu-H}
  (Z_\mu)^\nu=-\delta_\mu^\nu.
\end{equation}
%On the other hand, if $\partial H>0$ and there is a distinguished time
%coordinate in $\mathcal M$, then one can choose
%\begin{equation}\label{Zmunu+H}
%  (Z_\mu)^\nu=\eta_\mu^\nu\sqrt[n]{\partial H},
%\end{equation}
%where $\eta_\mu^\nu=\pm\delta_\mu^\nu$ with the plus sign for the
%distinguished time direction. Obviously, for the specific
%$p$-dependence in the proposition the square roots reduce to $1$.
It is clear that if $\partial H=0$ at same point the components
$(Z_\mu)^\nu$ of the vectorfields $Z_\mu$ cannot be linearly
independent $Z_\mu$ one from another and hence cannot span the $n$-dimensional 
tangent space on $\mathcal M$.

Comparing the next terms of $Y$ and $X$ one obtains
\begin{equation}
  \begin{split}
    {\scriptstyle\frac1{(n-1)!}}X^{A\nu_1\cdots\nu_{n-1}}
    \epsilon_{\nu_1\cdots\nu_{n-1}\rho}
    &=
    \sum_{\mu=1}^n(-)^{\mu+1}
    (Z_\mu)^A(Z_1)^{\nu_1}\cdots\widehat{(Z_\mu)^{\nu_\mu}}\cdots(Z_n)^{\nu_n}
    \epsilon_{\nu_1\cdots\widehat{\nu_\mu}\cdots\nu_{n}\rho}
    \\
    {\scriptstyle\frac1{(n-1)!}}{X^\sigma_A}^{\nu_1\cdots\nu_{n-1}}
    \epsilon_{\nu_1\cdots\nu_{n-1}\rho}
    &=
    \sum_{\mu=1}^n(-)^{\mu+1}
    (Z_\mu)^\sigma_A(Z_1)^{\nu_1}\cdots\widehat{(Z_\mu)^{\nu_\mu}}\cdots(Z_n)^{\nu_n}
    \epsilon_{\nu_1\cdots\widehat{\nu_\mu}\cdots\nu_{n}\rho}
  \end{split}
\end{equation}

%Let us return to the case when $\partial H$ vanishes at some point
%$\underline p\in\mathcal P$. Let $(Z_\mu)^\nu$ be a solution to (\ref{detZ})
%which does not vanish identically. By a coordinate transformation,
%$(Z_\mu)^n$, viewed as a matrix, can be brought into the Jordan normal
%form with $(Z_1)^1\neq 0$. Then one calculates at this point
%\begin{equation}
%  \begin{split}
%    (Z_1)^1\partial_{p^1_A}H
%    &=(Z_1)^\rho\partial_\rho^AH\\
%    &=(Z_1)^\rho{\scriptstyle\frac1{(n-1)!}}X^{A\nu_1\cdots\nu_{n-1}}_{(\underline p)}
%    \epsilon_{\nu_1\cdots\nu_{n-1}\rho}\\
%    &
%    =(Z_1)^\rho\sum_{\mu=1}^n(-)^{\mu+1}
%    (Z_\mu)^A(Z_1)^{\nu_1}\cdots\widehat{(Z_\mu)^{\nu_\mu}}\cdots(Z_n)^{\nu_n}
%    \epsilon_{\nu_1\cdots\widehat{\nu_\mu}\cdots\nu_{n}\rho}\\
%    &
%    =(Z_1)^A(Z_1)^\rho(Z_2)^{\nu_2}\cdots(Z_n)^{\nu_n}
%    \epsilon_{\nu_2\cdots\nu_n\rho}=0.
%  \end{split}
%\end{equation}
%Hence, in the transformed coordinates, $\partial_{p^1_A}H=0$ for all $A=1,\ldots,N$.

Now let $(Z_\mu)^\nu$ be given by (\ref{Zmunu-H}). Then
\begin{equation}\label{cond-decomposition}
  \begin{split}
    \partial_\rho^AH&={\scriptstyle\frac1{(n-1)!}}X^{A\nu_1\cdots\nu_{n-1}}
    \epsilon_{\nu_1\cdots\nu_{n-1}\rho}
    =
    (Z_\rho)^A\\
    -\partial_AH&=
    {\scriptstyle\frac1{(n-1)!}}{X^\rho_A}^{\nu_1\cdots\nu_{n-1}}
    \epsilon_{\nu_1\cdots\nu_{n-1}\rho}
    =(Z_\rho)^\rho_A,
  \end{split}
\end{equation}
which obviously is satisfied by 
\begin{equation}\label{soln-decomposition}
  \begin{split}
    (Z_\mu)^A&=\partial_\mu^AH\\
    (Z_\mu)^\nu_A&=
    -\frac1n\delta_\mu^\nu\partial_AH+(Z'_\mu)^\nu_A,.
  \end{split}
\end{equation}
where the $(Z'_\mu)^\nu_A$ are arbitrary functions that satisfy
\begin{equation*}
  (Z'_\mu)^\mu_A=0.
\end{equation*}
Note that the momentum directions of $Z_\mu$ are not given
uniquely. In particular, there are no conditions on the off-diagonal
terms $(Z_\mu)^\nu_A$, $\mu\neq\nu$. It remains to determine the components $(Z_\mu)_0$, but this
can be done using the third line in (\ref{HamVf-cond}). Indeed,
further comparison of (\ref{Y}) to (\ref{X}) yields
\begin{equation}
  \label{Zmu-0}
  -{\scriptstyle\frac{1}{(n-2)!}}X^{\sigma B \nu_1\cdots\nu_{n-2}}_A
  \epsilon_{\rho_1\nu_1\cdots\nu_{n-2}\rho_2}=
  (Z_{\rho_1})^B(Z_{\rho_2})^\sigma_A-
  (Z_{\rho_2})^B(Z_{\rho_1})^\sigma_A.
\end{equation}
Using (\ref{HamVf-cond}) we obtain for a special contraction 
\begin{equation}\label{delmuH}
  \begin{split}
    \partial_\mu H&=
    -{\scriptstyle\frac{1}{(n-2)!}}
    {X^\sigma_A}^{A\nu_1\cdots\nu_{n-2}}
    \epsilon_{\sigma\nu_1\cdots\nu_{n-2}\mu}
    -{\scriptstyle\frac{1}{(n-1)!}}
    X^{\nu_1\cdots\nu_{n-1}}_0\epsilon_{\nu_1\cdots\nu_{n-1}\mu}
    \\&=
    -\big((Z_\mu)^A(Z_\nu)^\nu_A-(Z_\nu)^A(Z_\mu)^\nu_A)\big)- (Z_\mu)_0.
  \end{split}
\end{equation}
This yields an expression for $(Z_\mu)_0$ in terms of the other
components of $Z_\mu$.

Equation (\ref{detZ}) shows that if $\partial H\neq 0$ then the
$(Z_\mu)$ are linearly independent (as their horizontal components
are). Hence $Y\neq 0$. Moreover, the components of $(Z_\mu)$ have been determined using
all of (\ref{HamVf-cond}). Thus  $Y$ is a Hamiltonean vectorfield
to $H$.\hspace*{\fill}$\Box$

\subsection{Solutions define decomposable Hamiltonean $n$-vectors}

As a next step, we ask what the Hamiltonean $0$-forms $H$ have to do
with the De Donder-Weyl Hamiltonean $\mathcal H$. Their relation is
already indicated in the notation of (\ref{goodH}) and can be guessed
further from (\ref{cond-decomposition}).
\begin{prop}
  Let $\gamma=(\varphi,\pi)$ be a solution of the De Donder-Weyl 
  equations (\ref{dDWHeq}) for some DW Hamiltonean $\mathcal H$.
  The tangent space of the image of $\gamma$ defines an $n$-vectorfield
  which is Hamiltonean with respect to the function $H$ given by
  (\ref{goodH}).  
\end{prop}
{\bf Remark.} From Lemma (\ref{decomposable}) in the appendix we know
that an $n$-vector $X$ is decomposable if and only if there are $n$
linearly independent vectors $Z_\mu$ which satisfy $Z_\mu\wedge
X=0$. This implies for the Hamiltonean $n$-vectorfields $X$ of the
given function $H$
\begin{equation}
  \label{motivation}
  0=(X\wedge Z_\mu)\PCPcontraction \Omega=Z_\mu\PCPcontraction dH.
\end{equation}
Combining (\ref{dDWHamiltonean}) and (\ref{Legendre}) we note that $H$
vanishes on sections $\gamma$ that satisfy the DW
equations. Therefore, it is natural to expect that $Z_1\wedge\cdots
\wedge Z_n$
is proportional to a Hamiltonean $n$-vectorfield $X$ if the
vectorfields $Z_\mu$ are lifts by $\gamma$.\\
{\bf Proof.}
In local coordinates, the section $\gamma$ is given by
\begin{equation}\label{Schnitt}
  \gamma(x)=(\varphi^A(x),\pi^\nu_A(x),-\mathcal H(x,\varphi(x),\pi(x)).
\end{equation}
Let $\partial_\mu$, $\mu=1,\ldots,n$, be a basis of $T_m\mathcal M$. Their
respective lifts $Z_\mu$ to $T\mathcal P$ via $\gamma$ are given by
\begin{equation}
  \begin{split}
    Z_\mu&=\partial_\mu+\partial_\mu\varphi^A\partial_A
    +\partial_\mu\pi^\nu_A\partial_\nu^A
    -[\partial_\mu \mathcal H+\partial_A \mathcal H\partial_\mu\varphi^A
    +\partial_\sigma^A\mathcal H \partial_\mu\pi^\sigma_A]\partial
    \\
    &=\partial_\mu+\partial_\mu^A \mathcal H\partial_A
    +\partial_\mu\pi^\nu_A\partial_\nu^A
    -[\partial_\mu \mathcal H+\partial_A \mathcal
    H\partial_\mu^A\mathcal H
    +\partial_\sigma^A\mathcal H \partial_\mu\pi^\sigma_A]\partial.
  \end{split}
\end{equation}
Note that the vectorfields $Z_\mu$ are not defined on all of
$\mathcal P$. Rather, they are given on the image of some local region
in $\mathcal M$ under
$\gamma$ only. 

Let $X$ be a Hamiltonean $n$-vectorfield and $\tilde Z_1\wedge \cdots\wedge
\tilde Z_n$ be a decomposition of it. Using the calculations of the
preceeding section, we conclude from equation (\ref{Zmunu-H})
\begin{equation}
  (\tilde Z_\mu)^\nu=-\delta^\nu_\mu=-(Z_\mu)^\nu,
\end{equation}
while from (\ref{soln-decomposition}) it follows that
\begin{equation}
  \begin{split}
    (\tilde Z_\mu)^A&=\partial_\mu^AH=-\partial_\mu^A\mathcal
    H=-(Z_\mu)^A\\
    (\tilde Z_\mu)^\mu_A&=-\partial_AH=\partial_A\mathcal
    H=-(Z_\mu)^\mu_A.
  \end{split}
\end{equation}
Finally, we compute for the remaining component $(Z_\mu)_0$
\begin{equation}
  \begin{split}
    (Z_\mu)_0&=-\partial_\mu\mathcal H
    -\partial_A\mathcal H\,\partial_\mu^A\mathcal H-
    \partial_\sigma^A\mathcal H\,\partial_\mu\pi^\sigma_A\\
    &=\partial_\mu H+\big((Z_\sigma)^\sigma_A\,(Z_\mu)^A
    -(Z_\sigma)^A\,(Z_\mu)^\sigma_A\big)
  \end{split}
\end{equation}
which goes over to (\ref{delmuH}) for $(\tilde
Z_\mu)_0=-(Z_\mu)_0$, $(\tilde Z_\mu)^A=-(Z_\mu)^A$, and 
$(\tilde Z_\mu)^\nu_A=-(Z_\mu)^\nu_A$. Therefore, the set of vectorfields
\begin{equation}
  \tilde Z_\mu=-Z_\mu,\quad \mu=1,\ldots,n
\end{equation}
defines a decomposition of a Hamiltonean $n$-vectorfield $X$ of $H$.
This proves the assertion. 
\hspace*{\fill}$\Box$

{\bf Remark.} At this point a remark is in order about the peculiar
form (\ref{goodH}). It is known that the De Donder-Weyl Hamiltonean 
(\ref{dDWHamiltonean}) constitutes a relation among coordinates of
$\mathcal P$ that describes the image of $\mathfrak F\mathcal L$. If
one wants to extract a function $\mathcal H_\Gamma$, the global
Hamiltonean function of \cite{EeMlRr}, out of it one needs to employ a
connection in $\mathcal E$,
\begin{equation}
  \mathcal H_\Gamma{(\textstyle x,v,\vec p)}=
  \mathcal H{(\textstyle x,v,\vec p)}
  -p^\mu_A \,\Gamma^A_\mu{(\textstyle x,v)}.
\end{equation}
Here we have used that every connection in $\mathcal E$ can be
interpreted as a map $\mathcal E\rightarrow\mathfrak J^1\mathcal E$.
Furthermore, with the help of the volume form $\omega$ on $\mathcal M$
for every connection $\Gamma$ there is a special function 
$\mathfrak p_\Gamma$
on $\mathcal P$ which uses that points in $\mathcal P$ are mappings of
the image of the connection $\Gamma$. In coordinates, 
\begin{equation}
  \mathfrak p_\Gamma{(\textstyle x,v,\vec p,p)}
  =p^\mu_A \,\Gamma^A_\mu{(\textstyle x,v)}+p.
\end{equation}
Combining these two, one obtains a function $H$ that is independent of
$\Gamma$,
\begin{equation}
  \begin{split}
    H{(\textstyle x,v,\vec p,p)}
    &=-\mathcal H_\Gamma{(\textstyle x,v,\vec p)}
    -\mathfrak p_\Gamma{(\textstyle x,v,\vec p,p)}\\
    &=-\mathcal H{(\textstyle
    x,v,\vec p)}-p.
\end{split}
\end{equation}

\subsection{Hamiltonean $n$-vectorfields on 
  $\tilde{\mathcal P}$\label{no-go}}

One might ask whether a Hamiltonean $n$-vectorfield on
$\tilde{\mathcal P}$ can be decomposable as well. We will show that this
is not the case for typical examples. For simplicity we shall assume
that the fibre bundle $\mathcal E$ admits a vanishing connection.\\
Again, we write a general ansatz for the $n$ vectorfields that shall
be combined to give a Hamiltonean $n$-vectorfield.
\begin{equation}\label{deftildeZmu}
  \tilde{Z}_\mu=\partial_\mu+(\tilde{Z}_\mu)^A\partial_A
  +(\tilde{Z}_\mu)^\nu_A\partial^A_\nu.
\end{equation}
An evaluation of the defining relation
\begin{equation}
  (\tilde{Z}_1\wedge\cdots\wedge\tilde{Z}_n)\PCPcontraction
  \Omega_\Gamma
  = d_\Gamma \tilde H
\end{equation}
for some function $\tilde H$
yields no condition on the $\partial_\mu$-components and the usual
ones on the terms containing one vertical vector, namely
\begin{equation}
  \begin{split}
    \partial_\rho^A\tilde H&=
    (\tilde Z_\rho)^A\\
    -\partial_A\tilde H&=
    (\tilde Z_\rho)^\rho_A.
  \end{split}
\end{equation}
Comparing this to the De Donder-Weyl equations (\ref{dDWHeq}) we
conclude that $\tilde H$ is to be interpreted as the De Donder-Weyl
Hamiltonean. \\
When looking at the $2$-vertical components one encounters a
difference, because the $dp\wedge d^nx$-term is missing in
$\Omega_\Gamma$. Therefore, instead of (\ref{delmuH}) one has
\begin{equation}
  \begin{split}
    0&=(\tilde Z_\mu)^A(\tilde Z_\nu)^\nu_A
    -(\tilde Z_\nu)^A(\tilde Z_\mu)^\nu_A\\
    &=
    -\partial_\mu^A\tilde H\,\partial_A \tilde H
    -\partial_\mu^A\tilde H\,(\tilde Z_\mu)^\nu_A.
  \end{split}
\end{equation}
Now let $\tilde H$ be given by
\begin{equation}
  \tilde H(x,v,\vec p)= {\textstyle\frac12}
  g_{\mu\nu}\eta^{AB} p^\mu_A p^\nu_B + V(x,v),
\end{equation}
where the function $V$ is arbitrary and $g$ and $\eta$ denote metrics
on space-time and the fibre, respectively. We now have
\begin{equation}
  0 =-g_{\mu\rho}\eta^{AB}p^\rho_B\,\partial_A V
    -g_{\nu\rho}\eta^{AB}p^\rho_B\,(\tilde Z_\mu)^\nu_A,
\end{equation}
from which by the independence of the polymomenta 
$p^\mu_A$ and the invertibility of $g$ and $\eta$ it follows that
\begin{equation}
  (\tilde Z_\mu)^\nu_A=-\delta^\mu_\nu \partial_A V.
\end{equation}
But this is in contradiction to $(\tilde Z_\mu)^\mu_A=-\partial_A V$
unless $n=1$ or $\partial_A V=0$.

\subsection{Integrability}

In the preceding subsections we have seen that Hamiltonean $0$-forms 
on $\mathcal P$ of the particular form 
\begin{equation}
  H(x,v,\vec p,p)=-\mathcal H(x,v,\vec p)-p,
\end{equation}
where $\mathcal H$ plays the r\^ole of the DW-Hamiltonean,
admit decomposable $n$-vectorfields which can be interpreted as
distributions on $\mathcal P$. The remaining question is whether there is an
integrable distribution among them. 
Of course, given a set of $n$ vectorfields that span the distribution
under consideration, by the theorem of Frobenius (\cite{KMS93}) one
just needs to verify that the vectorfields close under the Lie
bracket. However, as we have learned from (\ref{soln-decomposition}),
one cannot assign to a given Hamiltonean $0$-from $H$ a decomposition
$X_H=Z_1\wedge\cdots\wedge Z_n$ in a unique way. Rather, there is a
considerable arbitrariness in the choice of the polymomentum components
$(Z_\mu)^\nu_A$. This has to be fixed in a satisfactory way. In this
section, we will show that the required additional input comes from
solutions of the covariant Hamilton-Jacobi equations.

Let us first examine the case of
classical mechanics to understand the results below. 

In that case, to every time-dependent Hamiltonean there is a unique
(time-dependent) vectorfield on the doubly extended phase space. Of
course, this vectorfield can be integrated to yield a family of
integral curves. However, the vectorfield cannot in general be projected onto
the extended (covariant) configuration space $\mathbbm R\times \mathcal
Q$. Rather there is a correspondence between solutions
of the Hamilton--Jacobi equation and set of curves on $\mathbbm R\times
\mathcal Q$. 
More precisely, one is looking for a map $T$ that goes from $\mathbbm
R\times \mathcal Q$ to $\mathbbm R^2\times T^\ast\mathcal Q$ which
pulls back the Hamiltonean vectorfield onto the extended
configuration space. In the case of classical mechanics, this map
happens to be the gradient of another function $S$. 
For the curves thus obtained to be solutions to
the equations of motion, the function $S$ needs to satisfy an additional
equation, the celebrated Hamilton-Jacobi equation. In the simple case
of classical mechanics this procedure is somewhat superfluous as it
adds to the easy to handle set of ordinary differential equations a
partial differential equation, but in the general case $n>1$ this method
turns out to be quite helpful.

Let us come back to the case of a higher dimensional base manifold
$\mathcal M$. Here the fibre bundle $\mathcal E$ plays the r\^ole of
the extended configuration space, while the extended multisymplectic
phase space $\mathcal P$ replaces $\mathbbm R^2\times T^\ast\mathcal Q$. The
desired map $T:\mathcal E\rightarrow\mathcal P$, cf. the diagram
(\ref{Diagramm}), has to possess two 
properties. Firstly, there should be an integrable distribution on
$\mathcal E$ which is the pull back of some Hamiltonean $n$-vector
field to the given function $H$. Secondly, the integral manifolds have
to be solutions to the DW equations. Our aim will be to
give necessary and sufficient conditions  on $T$ for the resulting set
of integral submanifolds to be (local) foliations of $\mathcal
E$. This constitutes, of course, the best possible case, and for
general DW Hamiltoneans one will have to lower one's sights
considerably. In this article, however, we are aiming at some
geometrical picture and will, therefore, leave those matters aside.

\begin{prop}\label{integrable}
  Let $\mathcal H$ be a regular DW-Hamiltonean. Then one can find a local
  foliation of $\mathcal E$ where the leaves (when transported to
  $\mathcal P$ by virtue of the covariant Legendre map
  (\ref{Legendre}))   are solutions of the 
  DW equations if and only if there is a map 
  $T:\mathcal E\rightarrow \mathcal P$
  that satisfies in some coordinate system 
  \begin{align}
    \label{Bed-Integrabel}
    \partial_\mu^A\mathcal H(x,v,\vec T(x,v))&=0\\
    \label{Bed-Loesung-1}
    \partial_\mu T^\mu_A(x,v)&=-\partial_A\mathcal H(x,v,\vec T(x,v))\\
    \label{Bed-Loesung-2}
    \partial_\mu T_0(x,v)&=-\partial_\mu\mathcal H(x,v,\vec T(x,v))\\
    \label{Bed-Loesung-3}
    \partial_\mu T^\mu_A(x,v)&=-\partial_A T_0(x,v),
  \end{align}
  for all points $(x,v)$ in a local neighbourhood of 
  $\mathcal E$. 
  Here, $\vec T=(T^\mu_A)$ denotes the $p^\mu_A$-components of the
  map $T$ while $T_0$ stands for the value of the $p$-component of $T$.
\end{prop}
{\bf Remark.}
If the map $T$ can be written as a derivative with respect to the
field variables $v^A$ of a collection of functions $S^\mu$,
$\mu=1,\ldots,n$,
\begin{equation}\label{T=dS}
  T^\mu_A(x,v)=(\partial_AS^\mu)(x,v),\quad T_0(x,v)=(\partial_\mu S^\mu)(x,v)
\end{equation}
then the second set of equations,
(\ref{Bed-Loesung-1}) and (\ref{Bed-Loesung-2}), is a
consequence of the generalised Hamilton--Jacobi equation for the
functions $S^\mu$ (cf. \cite{Ru66}, ch. 4, sec. 2),
\begin{equation}\label{HamJacEq}
  \partial_\mu S^\mu(x,v)+\mathcal H(x,v,\partial_A S^\mu(x,v))=0.
\end{equation}
Clearly for $n=1$ the sum in the first term reduces to the (``time'')
derivative of some function $S$, and this equation becomes the
Hamilton--Jacobi equation. Note that the right hand side of the second equation of
(\ref{Bed-Loesung-1}) does not transform properly under a change
of coordinates. This corresponds to the fact that if one chooses a
different trivialisation, then the solutions to the DW equations will
not be constant anymore. In other words, the transformed map $T$ will
not satisfy the generalised Hamilton--Jacobi equations.\\
{\bf Proof of the proposition.}
Let $\mathcal U$ be an open subset of $\mathcal M$ such that there is a
local foliation of $\mathcal E$, i.e. a bijective map
\begin{equation}
  \varphi:\mathcal V\times\mathcal U\rightarrow \mathcal
  E\PCPrestrict{\mathcal U},
\end{equation}
where $\mathcal V$ denotes the typical fibre of $\mathcal E$. 
This defines a local trivialisation of $\mathcal E$ which will be used
for coordinate expressions for the rest of the proof. Furthermore,
one obtains a map $T:\mathcal E\rightarrow\mathfrak J^1\mathcal
E\rightarrow\mathcal P$
by taking the first jet prolongation of the section $\varphi(v,\cdot)$ for
every point $\varphi(v,x)$ and transporting (via the Legendre map)
this to $\mathcal P$. 
From 
\begin{equation*}
  (\partial_\mu^A\mathcal H)(x^\mu,v^A,(\partial_A^\mu
  L)(x^\mu,v^A,v^A_\mu))
  =v^A_\mu,
\end{equation*}
where $v^A_\mu$ gives the value of the derivative w.r.t. the
$\mu$-direction when evaluated on sections,
one concludes the first property. The remaining set of
equations then follows from the fact the the $\varphi(v,\cdot)$ are
solutions to the DW equations for every element $v\in
\mathcal V$ of the typical fibre.

Conversely, let $T$ be a map which fulfils the conditions of the
proposition. Then one can pull back a given decomposition of every
Hamiltonean vectorfield of $H$ to $\mathcal E$. Note that the
resulting vectorfields $\tilde{Z_\mu}$ are unique once the horizontal
component of 
the Hamiltonean $n$-vectorfield has been fixed as in (\ref{Zmunu-H}). From
(\ref{soln-decomposition}) one concludes that the resulting vector
fields are horizontal in the chosen coordinate system. Therefore, they
are integrable. Let 
\begin{equation}
  Z_\mu{(x,v,T^\nu_A(x,v),T_0(x,v))}
  =\partial_\mu+\partial_\mu
  T^\nu_A{\textstyle(x,v)}\,\partial_\nu^A
  -\partial_\mu \mathcal H{\textstyle(x,v,\vec p)}\,\partial,
\end{equation}
$\mu=1,\ldots,n$,be $n$ vectorfields on the image of $\mathcal E$
under $T$ ($T_0$ denotes the $p$-component of the map $T$). Then,
comparing the second set of conditions to the second set of equations
in (\ref{cond-decomposition}), it follows by virtue of
(\ref{Bed-Loesung-1}) and (\ref{Bed-Loesung-2}) that 
$Z_1\wedge\cdots\wedge Z_n$ is indeed a Hamiltonean 
$n$-vector to $H_{(x,v,\vec p,p)}=-\mathcal H_{(x,v,\vec p)}-p$. 
Furthermore,
as the tangent vectors $\widetilde{Z}_\mu$ on $\mathcal E$ do not have vertical
components in this coordinate system, their integral surfaces cannot
intersect. Hence, they describe a local foliation of $\mathcal E$.\\
Finally, having 
transported the sections from $\mathcal E$  via $T$ to $\mathcal
P$, their $p$-components 
by (\ref{Bed-Loesung-2}) and (\ref{Bed-Loesung-3})  can
differ from $-\mathcal H$ only by a constant.\\
\hspace*{\fill}$\Box$\\
{\bf Remark.} The extended multisymplectic phase space can be
identified with those $n$-forms on $\mathcal E$ that vanish upon
contraction with two vertical (w.r.t. the projection onto $\mathcal M$)
tangent vectors on $\mathcal E$. In coordinates, one has
\begin{equation}
  (x^\mu,v^A,p^\mu_A,p)\cong p^\mu_A\,dv^A\wedge d_\mu x + p\,d^nx.
\end{equation}
Hence, the map $T$ can be interpreted as an $n$-form on $\mathcal E$,
and equation (\ref{T=dS}) can be interpreted as 
\begin{equation}
  T=dS,
\end{equation}
while (\ref{HamJacEq}) becomes 
\begin{equation}
  H\circ dS =0.
\end{equation}
The conditions (\ref{Bed-Loesung-1})-(\ref{Bed-Loesung-3}) now can be
stated as
\begin{equation}
  d(H\circ T)=0,\quad dT=0.
\end{equation}

\subsection{An example: the free massive Klein-Gordon field}

To conclude this article we will give an example to show that the
assumptions of proposition \ref{integrable} are non-empty.\\
Let $L$ be the Lagrange function of the Klein--Gordon
field, i.e. let $\mathcal E$ be a trivial line bundle over $\mathcal
M=\Sigma\times\mathbbm R=\mathbbm R^4$ and
\begin{equation}
  L(x,v,v_\mu)=\frac12 g^{\mu\nu} v_\mu v_\nu-\frac12 m^2 v^2,
\end{equation}
where $g^{\mu\nu}$ denotes the metric tensor. The Euler-Lagrange
equation in this case is the celebrated Klein--Gordon equation
\begin{equation}\label{KGeq}
  (\Box+m^2)\Phi(\vec x,t)=0.
\end{equation}
As is well known, for every pair of functions $\varphi,\pi\in\mathcal
C^\infty(\Sigma)$
there is a unique function $\Phi\in\mathcal
C^\infty(\Sigma\times\mathbbm R)$
given by convolution with certain distributions $\Delta,\dot\Delta$,
\begin{equation}
  \Phi(\vec x,t)=(\Delta\ast\pi_0)(\vec x,t)+(\dot\Delta\ast\varphi_0)(\vec x,t),
\end{equation}
that satisfies the Klein--Gordon equation (\ref{KGeq}) and matches
with the initial data $\varphi,\pi$:
\begin{equation}
  \Phi(\vec x,0)=\varphi(\vec x),\quad (\partial_t\Phi)(\vec
  x,0)=\pi(\vec x).
\end{equation}
The corresponding DW Hamiltonean to $L$ is given by
\begin{equation}
  \mathcal H(x,v,p^\mu)=\frac12 g_{\mu\nu} p^\mu p^\nu+\frac12m^2v^2.
\end{equation}
Let $\varphi,\pi$ be a pair of initial data and $\Phi$ be the
corresponding solution.
The set of functions $S^\mu$ on $\mathcal E$ defined by
\begin{equation}
  S^\mu(x^\mu,v)=v g^{\mu\nu}(\partial_\mu\Phi)(x)
  -\frac 12\Phi(x) g^{\mu\nu}(\partial_\nu\Phi)(x).
\end{equation}
Clearly the $S^\mu$ satisfy
\begin{equation}
  \begin{split}
    (\partial_{p^\mu}\mathcal H)(x,\Phi(x),\partial_vS^\mu(x,\Phi(x)))
    &=g^{\mu\nu}(\partial_\nu\Phi)(x)
    \\
    (\partial_\mu S^\mu)(x,\Phi(x))
    &=-\mathcal H(x,\Phi(x),\partial_vS^\mu(x,\Phi(x))).
  \end{split}
\end{equation}
Therefore,
\begin{equation}
  \begin{split}
    X_\mu(x,v)&=\partial_\mu+\partial_\mu\Phi(x)\partial_v
    \\&\quad
    +\Big((\partial_\mu S^\nu)(x,\Phi(x))
    +(\partial_\mu\Phi)(x)(\partial_vS^\nu)(x,\Phi(x))\Big)
    \partial_{p^\nu}\\
    &\quad
    -\Big(\partial_\mu \mathcal H+\partial_v \mathcal H \,\partial_\mu\Phi(x)+
    \partial_{p^\nu}\mathcal H\,\partial_\mu S^\nu
    +\partial_{p^\nu} \mathcal H\,\partial_v S^\nu\,\partial_\mu\Phi(x)\Big)\partial_p
  \end{split}
\end{equation}
is a decomposition of a Hamiltonean $4$-vectorfield of 
$H(x,v,p^\mu,p)=-\mathcal H(x,v,p^\mu)-p$.

%%%%%%%%%%%%%%%%%%%%%%%%%%%%%%%%%%%%%%%%%%%%%%%%%%%%%%%%%%%%%%%%%%%%%%%%%%%%%%
\section{Conclusions}

We have clarified how $n$-dimensional submanifolds can be described by
decomposable $n$-fold antisymmetrised tensor products of
vectorfields. Those multivectorfields arise naturally in the context
of 
multisymplectic geometry, cf. equation (\ref{DefHamVf}). The
corresponding Hamiltonean forms are functions on the extended
multisymplectic phase space $\mathcal P$. If such a Hamiltonean function is of
the special form
\begin{equation}\label{special-form}
  H(x,v,\vec p,p)=-\mathcal H(x,v,\vec p)-p,
\end{equation}
then is admits a decomposable Hamiltonean $n$-vector by proposition
\ref{prop:decomposition}.\\
Conversely, if one is given a solution to
the DW equations with Hamiltonean $\mathcal H$, then its
associated multivectorfield is Hamiltonean for the function
(\ref{special-form}). The $p$-dependence characterises the orientation
of the solution submanifold as compared to the orientation on the base
manifold $\mathcal M$. Its origin can be understood in a geometrical way.\\
Thirdly, given a DW Hamiltonean function
(\ref{special-form}),  under certain additional conditions which
use a generalisation of the Hamilton--Jacobi theory of classical mechanics, one can
find an integrable Hamiltonean vectorfield on some subset of the
extended multisymplectic phase space. This multivectorfield foliates
the original fibre bundle where the 
theory has been formulated on. However -- in contrast to the case of
mechanics -- one does not have a unique local foliation of the extended
multisymplectic phase space $\mathcal P$ by solutions of the De
Donder-Weyl equations: Even for the mass free scalar wave equation one
can have two different solutions that coincide at one point with all
their first derivatives, i.e. polymomenta.\\
The question of integrability
does not arise in classical mechanics as there the equations of motion
are ordinary differential equations.\\

%%%%%%%%%%%%%%%%%%%%%%%%%%%%%%%%%%%%%%%%%%%%%%%%%%%%%%%%%%%%%%%%%%%%%%%%%%%%%%
\begin{ack}
  The authors  thank G. Barnich and I. Kanatchikov for fruitful
  discussions and M. Forger for pointing out reference \cite{Tul74}.
  C. P. thanks P. Bieliavsky for expressing his disbelief which has
  encouraged the completion of this work.
  We are grateful to H. A. Kastrup for drawing our attention to the
  comprehensive review \cite{Kas83} and to E. Goldblatt for mentioning
  \cite{EeMlRr}, where similar ideas have been developed. 
  In particular, the link between
  multivectorfields and distributions (in the sense of subspaces of
  the tangent space) can be found there in great detail. 
  Finally, we would like to thank
  I. Kanatchikov for his critical reading of the manuscript
\end{ack}

%%%%%%%%%%%%%%%%%%%%%%%%%%%%%%%%%%%%%%%%%%%%%%%%%%%%%%%%%%%%%%%%%%%%%%%%%%%%%%
\begin{appendix}

\section{Distributions and multivectors }

Much of this section seems to folklore by now. We add this material
for the sake of completeness. It can be found for instance in
\cite{EeMlRr,Martin:1988b}
Usually (\cite{KMS93}), when considering foliations of a given
manifold $\mathcal M$, one introduces the notion of distributions, i.e. the
determination of a sub vector space of $T\mathcal M$ at every point of
$\mathcal M$. Those sub vector space can be described by specifying a
basis at every point. This is somewhat ambiguous, but the
antisymmetrised tensor product of the chosen basis is unique up to a
pre-factor (the determinant of the basis transformation).
On the other hand, in multisymplectic geometry, the concept of Hamiltonean
$k$-vectors naturally arises, so it is plausible to examine the
correspondence of distributions and multivectors.
\begin{lem}\label{decomposable}
Let $V$ be an $n+m$-dimensional vector space over some field $\mathbbm{K}$ 
and $X$ an element of the $n$-th
antisymmetric tensor product of $V$, $X\in\Lambda^nV$. Then there are
$n$ linearly independent vectors $\{Y_i\}_{i=1,\ldots,n}$ that satisfy
\[
Y_i\wedge X=0
\]
if and only if 
\[
X=\lambda Y_1\wedge\cdots\wedge Y_n,
\]
where $\lambda$ is some element of $\mathbbm K$.
\end{lem}
For the proof, one chooses a basis of $V$ the contains the given
$Y_i$. Then every $n$-vector $X$ can be expanded in that basis, and
one can successively show that all components containing the extra
basis elements must vanish.\PCPqed
Obviously, there cannot be more than $n$ linearly independent vectors
annihilating $X$. For if there were, one would have 
\[
 0\neq Y_1\wedge\cdots Y_{n+1}=X\wedge Y_{n+1}=0,
\]
which is a contradiction.\\
There are, however, special cases, when the property of being
decomposable is always fulfilled apart from the trivial case
$X\in\Lambda^{\textrm{max}}V$. Namely, let $X$ be in $\Lambda^kV$ for
$k= \dim V-1$. Let $g(\cdot,\cdot)$ be a scalar product on $V$ and
$\ast$ be the corresponding Hodge star operation. Then, 
\begin{equation}
  \xi=\ast(X)\in V.
\end{equation}
Let $\eta_i$ be a basis of the orthogonal complement of
$\xi$. Obviously
\begin{equation}
  0=g(\xi,\eta_i)=\ast^{-1}(\eta_i\wedge\ast\xi)=\ast^{-1}(\eta_i\wedge X),
\end{equation}
hence $\eta_i\wedge X=0$. From the lemma, we conclude that $X$ is
the antisymmetrised tensor product of all $\eta_i$. This case
corresponds to the situation in $3$ dimensions. There, planes can be
described by $2$ linearly independent vectors (which is ambiguous) or
by indicating the vector perpendicular to the plane (which is unique
up to a pre-factor). The latter can be understood as the Hodge dual
(w.r.t. the scalar product that defines orthogonality) of the tensor
product of the former two.\\
On the other hand, let $V=\textrm{span}\{e_1,e_2,e_3,e_4\}$ and let 
$X=e_1\wedge e_2+e_3\wedge e_4$. One can easily check that indeed
there is no non-zero vector $v$ that annihilates $X$, i.e.
\begin{equation}
  X\wedge v=0\;\Leftrightarrow\;v=0.
\end{equation}

Now we are in the position to formulate in terms of multivectorfields
the condition of a
distribution $E$ on $\mathcal M$ to be integrable. A distribution is
integrable if every point of $\mathcal M$ belongs to some integral
manifold of $E$. Let the distribution $E$ be spanned by a set 
$\mathcal W$ of vectorfields on $\mathcal M$ at every point. 
Then (\cite{KMS93}, theorem 3.25) $E$ is integrable
if $\mathcal W$ is involutive, i.e. is closed under the Lie
bracket of vectorfields, and if $E$ is of constant rank along the flow lines
of all the vectorfields of $\mathcal W$. Conversely, the tangent
vectors of a given submanifold define local vectorfields that span a
distribution of constant rank and which are in involution. 
\begin{lem}
  Let $X_E$ be a multivectorfield that is associated with a
  $k$-dimensional distribution $E$ on some manifold $\mathcal M$. Then $E$ is
  integrable if and only if there are $k$ linearly independent local
  vectorfields
  $X_i$ that satisfy
  \begin{equation}
    [X_i,X]=\lambda_i X,\quad \lambda_i\in\mathcal C^\infty(\mathcal M),
  \end{equation}
  where $[\cdot,\cdot]$ denotes the Schouten bracket, which is a
  extension of the Lie bracket of vectorfields (\cite{Tul74}). For decomposable
  $n$-vectors, it is given by
  \begin{equation}\label{decomposable-schouten}
    \begin{split}
      [X,Y]&=\sum_{i=1}^p\sum_{j=1}^q(-)^{i+j}
      [X_i,Y_j]\wedge X_1\wedge\cdots \widehat{X_i}\cdots X_p\wedge
      Y_1\wedge\cdots \widehat{Y_j}\cdots Y_q.
    \end{split}
  \end{equation}
\end{lem}
{\bf Proof.} Using (\ref{decomposable-schouten}) one verifies that 
$[X_i,X]=\lambda_i X$ iff $[X_i,X_j]=f_{ij}^k X_k$, but the latter
condition means that the collection of all $X_i$ define a distribution
which is stable under the involutive closure of the $X_i$.\hspace*{\fill}$\Box$

\end{appendix}

%%%%%%%%%%%%%%%%%%%%%%%%%%%%%%%%%%%%%%%%%%%%%%%%%%%%%%%%%%%%%%%%%%%%%%%%%%%%%%


\begin{thebibliography}{99}
\bibitem {CantrijnIbortDeLeon:1999}
  {J. F. Cari\~nena, M. Crampin, A. Ibort:}
  {\em On the multisymplectic formalism for first order field theories.}
  Differ. Geom. Appl. {\bf 1} (1991), 345-374.

\bibitem {Dic91} L. A. Dickey: 
  {\em Soliton Equations and Hamiltonian Systems.}
  World Scientific, Singapore, 1991.

\bibitem {Di90}
  J. Dito: 
  {\em Star-Product Approach to Quantum Field Theory: The Free Scalar Field.}
  {Lett. Math. Phys.} {\bf 20} (1990), 125--134.

\bibitem {DodsonPoston:1979}
  {C. T. J. Dodson, T. Poston:}
  {\em Tensor Geometry}, Pitman, London, 1979.

\bibitem {DuFr00}
  M. D\"utsch, K. Fredenhagen:
  {\em Algebraic Quantum Field Theory, Perturbation Theory, and the Loop
    Expansion}, {\tt hep-th/0001129}.

\bibitem {EeMlRr}
  A. Echeverr\'{\i}a-Enr\'{\i}quez, M. C. Munoz-Lecanda, N.  Roman-Roy:
  {\em Multivector fields and connections: setting Lagrangian equations in
  field theories.}
  {J. Math. Phys.} {\bf  39}, No.9 (1998), 4578-4603, 
  {\tt dg-ga/9707001}.

\bibitem {ForgerRoemer:2000}
  M. Forger, H. R\"omer: 
  {\em A Poisson Bracket on Multisymplectic Phase Space.}, {\tt
    math-ph/0009037}, to appear in the proceedings of the 32nd
  Symposium on Mathematical Physics, Toru\'n, Poland, June 2000.

\bibitem {GoIsMa98}
  M.~J. Gotay, J. Isenberg, J.~E. Marsden: 
  {\em Momentum Maps and Classical Relativistic Fields I: Covariant
    Field Theory\/}, 
  {\tt physics/9801019}.

\bibitem {Greub:1978}
  {W. Greub:} {\em Multilinear Algebra}, 
  2nd ed., Springer Verlag, N.Y.,
  1978.

\bibitem {Kanatchikov:1997}
  I.~V. Kanatchikov: 
  {\em On field theoretic generalizations of a {P}oisson algebra.}
  {Rep. on Math. Phys.}  {\bf 40}.2 (1997), 225--234,
  {\tt hep-th/9710069}.

\bibitem {Kanatchikov:1998}
  I.~V. Kanatchikov: 
  {\em Canonical structure of classical field theory in
  the polymomentum phase space.}
  {Rep. on Math. Phys.}  {\bf 41}.1 (1998), 49--90,
  {\tt hep-th/9709229}.
\bibitem {Kas83}
  H. A. Kastrup: 
  {\em Canonical Theories of Lagrangian Dynamical Systems in
  Physics.}
  {Physics Reports }{\bf 101} (1983), 1--167.

\bibitem {KMS93}
  I. Kol{\'{a}}{\v{r}}, P.~W. Michor, J. Slov{\'{a}}k:
  {\em  Natural Operations in Differential Geometry}.
  Springer-Verlag, Berlin, Heidelberg, New York, 1993.

\bibitem {Martin:1988a}
  G. Martin:
  {\em A Darboux Theorem for Multi-Symplectic Manifolds.}
  Lett. Math. Phys. {\bf 16} (1988), 133-138.

\bibitem {Martin:1988b}
  G. Martin:
  {\em Dynamical Structure for k-Vector Fields.}
  Intern. J. Th. Phys. {\bf 27}, No. 5 (1988), 571-585.

\bibitem {Noe18b}
  E. Noether: 
  {\em Invarianten beliebiger {D}ifferentialausdr{\"u}cke.}
  {Nachr. Kgl. Ges. Wiss. G{\"o}ttingen, Math.-phys. Kl.} 
  (1918), 37-44.

\bibitem {Pa00}
  C. Pauf\/ler: {\em A Vertical Exterior Derivative
    in Multisymplectic Geometry and a Graded Poisson Bracket for 
    Nontrivial Geometries},
  {\tt math-ph/0002032}, 
  Rep. on Math. Phys. {\bf 47} (2001), 101-119.\\
  C. Paufler: {\em On the Geometry of Field Theoretic Gerstenhaber
    Structures.} {\tt math-ph/0102012}, to appear in the proceedings
  of the 32nd Symposium on Mathematical Physics, Toru\'n, Poland, June 2000.

\bibitem {RoPa99}
  H. R\"omer, C. Pauf\/ler: {\em Anomalies and Star Products.}, in
  Quantum Theory and Symmetries, H.-D. Doebner, J.-D. Hennig,
  W. L\"ucke, and V. K. Dobrev eds., World Scientific, Singapore,
  2000,   {\tt hep-th/0010067}.

\bibitem {Ru66}
  H. Rund:
  {\em The {H}amilton-{J}acobi Theory in the Calculus  of Variations.}
  \newblock Hazell, Watson and Viney Ltd., Aylesbury, Buckinghamshire, U.K.,
  1966.

\bibitem {Sau89}
  D.~J. Saunders: {\em The Geometry of Jet Bundles}.
  \newblock {\em Lond. Math. Soc. Lect. Notes Ser., 142}.
  \newblock Cambr. Univ. Pr., Cambridge, 1989.

\bibitem {Trau67}
  A. Trautman:
  {\em Noether Equations and Conservation Laws}.
  {Commun. Math. Phys.}  {\bf 6} (1967), 248--261.
  
\bibitem{Tul74}
  W. M. Tulczyjew:
  {\em The graded Lie algebra of multivector fields and the generalized Lie
  derivative of forms. }
 \newblock Bull. Acad. Pol. Sci., Ser. Sci. Math. Astron. Phys. 
 {\bf 22} (1974), 937--942.
\end{thebibliography}
\end{document}